\documentstyle[12pt,fleqn]{article}
\newcommand{\be}{\begin{equation}}
\newcommand{\ee}{\end{equation}} 
\newcommand{\n}[1]{\label{#1}}

    \newcommand{\ha}{\frac{1}{2}}

\newcommand{\noi}{\noindent}

\newcommand{\ga}{\alpha}

\newcommand{\gc}{\gamma}
\newcommand{\gd}{\delta}

\newcommand{\gth}{\theta}

\newcommand{\gl}{\lambda}

\newcommand{\gs}{\sigma}

\newcommand{\gff}{\varphi}

\newcommand{\gD}{\Delta}

\newcommand{\gF}{\Phi}

\newcommand{\gW}{\Omega}

\newcommand{\ra}{\rightarrow}

\newcommand{\laar}{\longleftrightarrow}

\newcommand{\ov}{\overline}

\newcommand{\lan}{\langle}
\newcommand{\ran}{\rangle}

\title{The Topology of The Cosmic Microwave Background Anisotropy  on The Scale
$\sim1^o$}
\author{D.I.\ Novikov$^{1,2,3^*}$ and H.\ E.\ J\o rgensen$^{1^*}$ \\
\small
1) University Observatory, {\O}ster Voldgade 3, DK-1350 Copenhagen K, Denmark\\
\small
2) Theoretical Astrophysics Center, Blegdamsvej 17, DK-2100, Copenhagen \O,
Denmark\\
\small
3)  Astro Space Center of P.N.Lebedev Physical Institute\\
\small
Profsoyuznaya 84/32, Moscow, 117810, Russia\\
-----------------------------------------------\\
\small
*) Permanent address}
\normalsize

\date{}

\begin{document}
\maketitle

\begin{abstract}

In this paper we develop the theory of clusterization of peaks
in a Gaussian random field. We have obtained new mathematical
results from this theory and the theory of percolation and have proposed
a topological method of analysis of sky maps based on these
results. We have simulated $10^o\times10^o$ sky maps of the cosmic microwave
background anisotropy expected from different cosmological models
with $0.5^o-1^o$ resolution in order to demonstrate how this method
can be used for detection of non-Gaussian noise in the maps and
detection of the Doppler-peak in the spectrum of perturbation of
$\gD T/T$.

\end{abstract}
PACS number(s) 98.80.Cq, 97.60.Lf, 98.70.Vc
\section{Introduction}
Observations of the cosmic microwave background (CMB) anisotropy are
fundamental in understanding the formation of structures and the
nature of the dark matter in the Universe. The distribution of the
fluctuations of $\gD T$ on angular scales larger than a few degrees
characterizes fluctuations of the primordial density on scales
larger than the acoustic horizon at the moment of the last scattering.
In this paper we discuss the CMB fluctuations on the angular
scale $\gth\sim 1^o$. This angular size corresponds to the scale at
recombination. The distribution of the CMB anisotropy on this scale
preserves the information about the ionization history of the Universe and
about the nature of the dark matter. Various scenarios of recombination
and different amounts of baryons in  different cosmological models
lead to different properties of the anomaly in the $\gD T$ distribution.
(Namely, the different height of the so-called Doppler-peak in the power
spectrum $C_l$ at the multipole number $l\approx 220$). (Efstathiou,
Bond, White, 1992; Coulson et al., 1994; Gorski, 1993; Scott et al., 1995).

Several groups have reported on observational data on
angular scales about $1^o$ (Devlin et al., 1994; Clapp et al., 1994;
Dragovan et al., 1994; De Bernardis et al., 1994; Cheng et al., 1994;
Schuster et al., 1993; Wollack et al., 1994). The interpretations of
these experimental results and the comparison  with the expected power
spectrum of the $\gD T/T$ fluctuations in different cosmological models are
unclear.

Several authors propose different methods to detect the peak-like
anomaly in the angular distribution. These methods are based on
correlation analysis (Hinshaw et al., 1995; Naselsky and I.Novikov
1993, J\o rgensen et al., 1995). Recently, the first attempt to
understand the topological properties of $\gD T/T$ maps caused by the
presence of the Doppler-peak, was made by Naselsky and D.\ Novikov
(1995).  Many important questions related to this problem were
listed in the paper by Hinshaw, Bennett and Kogut (1994).  Some of them
are: 1) how to distinguish the effect caused by the presence of the
Doppler-peak from the isolated point sources, and 2) what is the role of
the cosmic variance in a statistical analysis of small regions of the
sky.

In this paper, we propose the topological method of analysis of the sky
maps. This method is based on the assumption that the initial
fluctuations are Gaussian.  In this case, the background radiation will
form a two-dimensional Gaussian random field. The statistical properties
of a  Gaussian random field were first investigated by Rice (1944,1945)
for a one-dimensional field to analyze electrical noise in communication
devices.  A.Doroshkevich (1970) was the first who applied this theory
extensively in the study of formation of cosmic structures.
J.Bardeen J.Bond, N.Kaiser and A.Szalay (1986) (referred to as BBKS in
what follows) in their classical paper developed the theory for a
three-dimensional field and J.Bond and G.Efstathiou (1987) (referred to
as BE in what follows) developed the theory for a two-dimensional field.

In the work presented here, we develop the theory of clusterisation of
peaks in a random Gaussian field, investigate the influence of the
spectral parameters on the $\gD T/T$ fluctuations for different
cosmological models, propose methods of filtering of the sky maps,
investigate the properties of the two-point correlation function for a
small region of the sky, investigate the topological properties of the
$\gD T/T$ angular distribution caused by the presence of the
Doppler-peak in the spectrum of fluctuations of $\gD T/T$, and propose
the method of percolation and cluster analysis which allows us to
detect the unresolved point sources (non-Gaussian noise which can
closely imitate the presence of a Doppler-peak in the spectrum) and
remove them from the observational data.

Note, that percolation has become a popular term among astronomers and
cosmologists. It was first introduced in cosmology by Zel'dovich (1982)
and has been succesfully applied during several years  for the investigation
of density distributions in the non-linear stage of gravitational
instability. (See the review by Dominik and Shandarin, 1992; Einasto et
al., 1984; Klypin, 1985). The $\gD T/T$ maps percolation as a first
test of the presence of non-Gaussian noise was first discussed by
Naselsky and D.Novikov (1995).

The outline of this paper is as follows. In Section 2, we  review some
general properties of $\gD T/T$ fluctuations and investigate the
properties of the correlation function  for different cosmological
models, namely with and without a Doppler-peak. In Section 3, we
introduce different kinds of filtering of the $\gD T/T$ maps and
discuss the influence of filtering on spectral parameters and the
correlation function. In Section 4, we develop the properties of
one-dimensional slices of a two-dimensional Gaussian random field and
of a two-dimensional Gaussian random field itself and obtain what we
consider to be the most important result of this paper: that the
clusterisation of peaks in a Gaussian random field depends only on one
spectral parameter $\gc$ and is determined by the Gaussian nature of the
field. In the same section we also propose a method for detecting
``noise peaks''. In Section 5, we propose a method of analysis of the
sky maps. We summarize our main results in Section 6. Details of the
derivations may be found in the appendices.

\section{The power spectrum  and correlation function on intermediate
scales}

In this paper we assume that the fluctuations in the cosmic microwave
background (CMB) are the result of a random Gaussian process. This
hypothesis has been adopted by many authors (see the review by
B.E.Hinshaw et al., 1995) and may be argued for in the following way:
since the fluctuations of the microwave background were imprinted in
the linear regime, $\gD T/T$ will be a linear combination of the
initial perturbations amplitudes, and will, therefore, have the same
probability distribution. Such a field with a zero mean is completely
characterized by its two-point correlation function or by the
equivalent power spectrum.

Let us consider the distribution of the temperature of the cosmic
microwave background on the celestial sphere. As it was mentioned
above, we assume that the field is a two-dimensional random Gaussian
field on a sphere. This field is completely characterized by the
power spectrum $C_l$. Using this description one can write the
following well known expression for the temperature of the relic
radiation
\be
T(\ov{q})=\lan T(\ov{q})\ran+\sum^{\infty}_{l=1}\sum^l_{m=-l}
a^m_lC_l^{\frac{1}{2}}Y^m_l(\ov{q}),  \n{2.1}
\ee
where $\ov{q}$ is the unit vector tangential to the direction of photon
motion; $a^m_l$  are independent random Gaussian numbers; $\lan
T(\ov{q})\ran$ is the average temperature of the relic radiation, so that
$\lan T(\ov{q})\ran=\frac{1}{4\pi}\int T(\ov{q})d\gW$;\
$Y^m_l$ are the spherical harmonics.

We introduce the following expression for the anisotropy of the CMB:
$\gD T(\ov{q})=(T(\ov{q})-\lan T(\ov{q})\ran)/\lan T(\ov{q})\ran$. The
two-point correlation function $C(\gth)$ can be found  by averaging
of $\gD T(\ov{q})\cdot \gD T(\ov{q}^{\prime})$ over the whole sky under
the condition that the angle between the directions of $\ov{q}$ and
$\ov{q}^{\prime}$ is a constant:
\be
C_{obs}(\gth)=\lan\gD T(\ov{q})\cdot \gD T(\ov{q}^{\prime})\ran,\ \ \
\ov{q}\cdot\ov{q}^{\prime}=\cos\gth\ .   \n{2.2}
\ee
Taking into account Eq.(2.1) and that $\lan a_l^ma_{l^{\prime}}^{m^{\prime}}
\ran =\gd _{ll^{\prime}}\gd_{mm^{\prime}}$ we get:
\be
C_{obs}(\gth)=\frac{1}{4\pi}\sum_{l=2}^{\infty}
\sum_{m=-l}^l(a_l^m)^2C_lP_l(\cos\gth).   \n{2.3}
\ee
The average value of the observable correlation function is
\be
C(\gth)= \ov{C_{obs}(\gth)}=
\frac{1}{4\pi}\sum_{l=2}^{\infty}(2l+1)C_lP_l(\cos\gth)\ . \n{2.4}
\ee
We begin the summation in  Eq.(2.4) from $l=2$.  The term with $l=1$
has to be removed before the calculation of the correlation function
because the contribution to this term in the $\gD T/T$ fluctuations
cannot be separated from  the motion of the observer relative to the
background radiation.

Note, that due to the presence  of the Doppler-peak in the spectrum of
perturbations of $\gD T/T$,  the two-point correlation function for the
CMBR anisotropy has a very characteristic feature on the angular scale
of the last scattering, $\theta\sim \theta_{ls}$.  This characteristic
feature looks like a local antipeak on this scale.

In practise the Doppler-peak in the spectrum of perturbations leads to
an effective decrease of the correlation radius. On Fig.1 we have
plotted the two-point correlation function for three different models
($\gW_b=0.03$, $h=0.5$;  $\gW_b=0.1$, $h=0.5$; and the model without a
Doppler-peak) in order to illustrate their properties on a scale
$\gth\sim 1^o$.

Let us consider the influence of a finite size of the considered
region on the property of the observed correlation function. It is obvious,
that the observed correlation function differs from its average
ensemble approximately by the value of the dispersion
\be
D_o(\gth)=\ov{C^2_{obs}}(\gth)-\ov{C_{obs}}^2=
\left(\frac{1}{4\pi}\right)^2\sum_l(2l+1)C^2_lP^2_l(\cos\gth)\ . \n{2.5}
\ee
The subscript $0$ on the left-hand side  means that this value is obtained
by averaging over the whole sky. The value $D_0(\gth)$ is very small for
$\gth\sim 1^o$, but if we consider only a small part of the sky, then this
value increases to:
\be
D_{\gW}(\gth)\sim\sqrt{\frac{4\pi}{\gW}}D_0(\gth), \n{2.6}
\ee
where $\gW$ is the angular size of the considered region. On Fig.2 we
have plotted the correlation function for $\gW_b=0.03$, its dispersion
and realizations of the random process for a part of the
sky of $10^o\times10^o$ for the unfiltered and filtered cases presented on
Maps 1a and 1b. This result demonstrates  the difficulty to detect
the Doppler-peak using only small regions of the sky.

Note, that this difficulty is caused only by ``cosmic variance''. If
the presence of the noise is taken into account, then this difficulty
becomes even  worse (see Hinshaw et al., 1995). Further we are
interested in detection of the peak by using not only correlation
analysis of the observational data, but also by topological analysis of
the sky maps. For the further investigations we introduce the filtering
of sky maps.

\section{Filtering of the sky maps}

\begin{center}
\noi{\bf a. Filtered correlation functions}
\end{center}

If we only consider a small part of the sky, then the geometry is
approximately flat, and we can introduce the Dekart coordinates on a
small area of a unit radius sphere ($\gth\times\gth$, $\gth\ll\pi$)  and
describe $\gD T(x,y)$ as a sum of the Fourier series  (B.E.):
\be
\gD T(x,y)= \sum_{ij}a_{ij}C^{\frac{1}{2}}(k)\cos
\left(2\pi\frac{ix+jy}{L}+\gff_{ij}\right),  \n{3.7}
\ee
where $C(k)$ is the power spectrum; $k=\frac{2\pi}{L}\sqrt{i^2+j^2}$;
$a_{ij}$  are independent random Gaussian values; $\gff_{ij}$ are
random phases homogeneously distributed in the interval $(0,2\pi)$;
$L\approx \gth_0$ is the size of the investigated area.

Note, that Eq.(3.7) takes into account only modes smaller than
$L$ and cannot be applied for simulation of the map $L\times L$.
Nevertheless, after simulation of the map $L\times L$ by (3.7), it is
possible to use only the part of this map $l\times l$, where $l\ll L$
since the  simulated distribution of $\gD T$ in this part contains all modes
and can be used for  imitation of the $\gD T$ distribution on the celestial
sphere.

The correlation function $C_{obs}(r)=\lan\gD T(x,y)\cdot\gD
T(x^{\prime},y^{\prime})\ran$ can be found by averaging over the square
$L\times L $ similar to Eq.(2.2).
\be
C_{obs}(r)=\frac{1}{2}\sum_{ij}a^2_{ij}C(k)J_0(kr), \n{3.8}
\ee
where $r=\sqrt{(x-x^{\prime})^2+(y-y^{\prime})^2}$.  After averaging over the
ensemble, we have
\be
C(r)=\ov{C_{obs}(r)}=\frac{1}{2}\sum_{ij}C(k)J_0(kr).  \n{3.9}
\ee
where $k$ was given above as function of $i$ and $j$.
Eq.(3.9) is in good agreement with  Eq.(2.4) because  if
$\gth\ll\pi$ and $l\gg 1$ then $P_l(\cos\gth)\approx J_0(l\gth)$, and
$l\gth\approx kr$.

Thus, the fluctuations of $\gD T$ can be described by Eq.(3.7), where
$C(k)\approx C_l$, $k\sim l/ \xi_n$ ($\xi_n$ is the present horizon),
and $C_l$ are the well known coefficients of the correlation function
decomposition to  series of Legendre polinomials, for different models
of the Universe.  It is obvious that the observational correlation
function differs from its average over the ensemble  by the value of the
dispersion:
\be
D(r)=\ov{C^2_{obs}(r)}-(\ov{C_{obs}(r)}
)^2=\frac{1}{2}\sum_{ij}C^2(k)J_0^2(kr)\ . \n{3.10}
\ee
For $(r\ll L)$ the correlation function and its dispersion can be written
as follows:
\be
\begin{array}{lc}
C(r)=\pi\int kC(k)J_0(k r)dk\\
D(r)=\pi\int kC^2(k)J^2_0(kr)dk\\
C_{obs}(r)\sim C(r)\pm\sqrt{D(r)}. \n{3.11}
\end{array}
\ee

Note that for investigation of scales as the size of the
recombination horizon, $R=R_{LS}$,
the size of area has to be $\gth_{LS}\ll l\approx\gth_0\ll
\pi$, $\gth_{LS}\simeq\frac{R_{LS}}{\xi_n}$

For  the further discussion we
introduce smoothing of the map by a
Gaussian filter:
\be
H_0(\gd,r)=\frac{1}{4\pi\gd^2}e^{-\frac{r^2}{4\gd^2}}; \hspace{1cm}
r^2=(x'-x)^2+(y'-y)^2\ . \n{3.12}
\ee
Then the filtered field $\gD\widetilde{T}(x,y)$ is calculated from the
initial, i.e., observable field $\gD T(x,y)$ by:
\be
\gD\widetilde{T}(x,y)=\frac{1}{4\pi\gd^2}\int_{-\infty}^{+\infty}
\int_{-\infty}^{+\infty} \gD T(x',y')e^{-\frac{r^2}{4\gd^2}}dx'dy'\ .
\n{3.13}
\ee
Formally, the limits of integration in Eq.(3.13) are
$(-l/2,l/2)$, but, if $\gd\ll L$, we can use $\pm\infty$ instead of
$(-l/2,l/2)$ as the limits of integration.  Thus, the
correlation function for the filtered map is:
\be
C(r,\gd)=\lan\gD\widetilde{T}(x,y)\cdot\gD\widetilde{T}(x',y')\ran=
\pi\int_0^{\infty}kC(k)e^{-\frac{k^2\gd^2}{2}}J_0(k r)dk\ . \n{3.14}
\ee
Another kind of filtering is the difference between two Gaussian
filters:
\be
H_*(\gd_{min},\gd_{max},r)=H_0(\gd_{min},r)-H_0(\gd_{max},r)\ .
\n{3.15}
\ee
This filter removes the influence of the long modes. The spectrum of
multipoles $C_l$ for many popular models has a specific shape:
\be
C_l\sim \frac{1}{l(l+1)}\ .\n{3.16}
\ee
Therefore, it is convenient to introduce a filter $H_2(\gd,r)$ which
has the following characteristic feature:  the spectrum of the filtered
map, $\widetilde{C}(k)$, can be obtained from the spectrum of
unfiltered map $C(k)$ by:
\be
\widetilde{C}(k)=k^2C(k)e^{-\frac{k^2\gd^2}{2}}\ . \n{3.17}
\ee
If $C(k)=\frac{1}{k^2}$ (spectrum without Doppler-peak), then from
Eq.(3.14) and Eq.(3.17) we can easily see, that the correlation function for
the filtered map has an especially simple (Gaussian) form:
\be
\frac{C(r,\gd)}{C(0,\gd)}=e^{-\frac{r^2}{2\gd^2}}\ . \n{3.18}
\ee

The deviation from the Gaussian in the observed correlation function
for a filtered map can be caused either by  presence of the Doppler-peak
or by ``cosmic variance'' and noise. On Fig.1(a,b,c) we have plotted filtered
correlation functions with different kinds of filters. This filter also
improves the statistical properties of the map, which is
convenient  for the further cluster analysis. On the Map.1a,b we have
shown the $\gD T$ distribution after filtering and unfiltered, respectively.

Now let us consider, how the filtering influences the spectral
parameters.

\begin{center}
\noi{\bf b. Spectral parameters}
\end{center}

Using, Eq.(3.11), we can introduce  the spectral parameters similarly to
(BBKS):

\be
\begin{array}{lc}
\gs_0^2=\pi\int kC(k)dk\\
\gs_1^2=\pi\int k^3C(k)dk\\
\gs_2^2=\pi\int k^5C(k)dk\\
R_*=\frac{\gs_1}{\gs_2},\hspace{1cm} \gamma=\frac{\gs_1^2}{\gs_0\gs_2}\ .
\end{array}
\n{3.19}
\ee
These parameters are completely defined by the value of correlation
function and its second and fourth derivatives at the zero.

In the next section we will demonstrate, that the value of $\gc$
determines the geometry of the $\gD T$ distribution seen on the map. In
Fig.3a,b we have plotted the dependence of $\gc$ on the value of resolution
for different kinds of filter. From this figure we can easily see that
the parameter $\gc$, for the model with a Doppler-peak, is different
from $\gc$ for the model without a peak for an appropriate value of
$\gd$ (for a resolution angle which corresponds to $l\sim 150$). From Fig.1
we also see that the differences between the correlation functions for the
different models is particularly evident for this specific value of $\gd$.
We name this value as the resonance filter value, $\gd_{res}$.

\section {Clusterisation of peaks in a random Gaussian field}

A Gaussian random field is the field for which a joint Gaussian probability
distribution for random variables $x_i$ is:
\[
P(x_1, ...., x_n) dx_1 .... dx_n=\frac{e^{-Q}}{((2\pi)^n \det M)^{1/2}}
dx_1 ... dx_n,
\]
\be
2Q=\sum_{ij}\gD x_i(M^{-1})_{ij}\gD x_j\ . \n{4.20}
\ee
Only the means of the random variables $\lan x_i\ran$ and their
variances are required to specify completely the covariance matrix $M$
and the distribution
\be
M_{ij}=\lan\gD x_i\gD x_j\ran, \hspace{1cm} \gD x_i=x_i-\lan x_i\ran. \n{4.21}
\ee

\subsection{One-dimensional cross-section of the two-dimensional field}

The problem of the clusterisation of peaks is especially easy in
one-dimension. We demonstrate it in order to illustrate the properties
of clusters of maxima in one dimension.  Note, that results of this
section can be applied to the analysis of one-dimensional observation.
We now introduce the definition of a cluster of maxima of $\gD T(x)$.
Let us consider two points, $x_1$ and $x_2$, $(x_1<x_2)$ and a threshold
level $\nu_t$ (where $\nu=\frac{\gD T}{\gs_0})$, such that $\gD T(x)>\nu_t$ for
all $x$ in the
interval $[x_1;x_2]$. We call these two points connected to each other
with respect to the threshold $\nu_t$.  A collection of $k$ maxima (of $\gD
T(x)$) located at the points $x_1, x_2, ..., x_k$ is called a cluster
of $k$ maxima if all the points $x_1, x_2, ..., x_k$ are connected to
each other with respect to the threshold level
$\nu_t$:
\be
\gD T(x)>\nu_t, \hspace{1cm} x_1\le x\le x_k\ .\n{4.22}
\ee
In what follows, we will estimate the probability of
existence of such a configuration.

\begin{center}
{\bf a. One-dimensional distribution}
\end{center}

If we consider a one-dimensional slice of the two-dimensional field, then the
distribution of $\gD T(x)$ can be written using Eq.(3.7)
\[
\gD T(x)=\sum_{ij}a_{ij}\sqrt{C(k)}\cos\left(\frac{2\pi}{L}ix+\gff_{ij}\right)
\hspace{1cm}k=\frac{2\pi}{L}\sqrt{i^2+j^2}\ .
\]

The two-point correlation function in this case is:
\[
C(r)=\lan\gD T(x) \gD T(x')\ran=\pi\int kC(k)J_0(kr)dk,
\hspace{1cm} r= |x-x'|,
\]
similar to the two-dimensional case.
Next, we introduce new variables:
\be
\nu(x)=\frac{\gD T(x)}{\gs_0};\hspace{1cm} \eta(x)=\frac{\gD T'(x)}{\gs_1};
\hspace{1cm} \xi(x)=\frac{\gD T''(x)}{\gs_2}\ .  \n{4.23}
\ee
\begin{center}
{\bf b. Conditional probability}
\end{center}

We shall estimate the probability that two given maxima at the points
$x_1$ and $x_2$ are connected together. A way to solve
this problem is to find the total num of minima  above some level
$\nu_t$, $n^+_{min}(r)$, and the total number of all minima,
$n_{min}(r)$, between these two points. The ratio of these two values
is an estimate of the desired probability.

We introduce the following notation. The event that the minimum  with
value above $\nu_t$ is located at the point $x$ is called the event
$A$.  The event that the maximum  with value above $\nu_t$ is located
at the point $x_1$ is called the event $B$. The event that such a
maximum is located at the point $x_2$ is called the event $C$.  Then
the conditional probability that minimum at the point $x$ between two
given maxima is above some threshold $\nu_t$ can be calculated via the
Bayes formula:
\be
P(A|B;C)=\frac{P(A;B;C)}{P(B;C)}\ . \n{4.24}
\ee
The obvious joint probabilities
\[
P(A;B;C)=\left(\gD x
\right)^3 \int|\xi_1\xi_2\xi|P(\nu_1,\eta_1=0,\xi_1, \nu_2,\eta_2=0,\xi_2,
\nu,\eta=0,\xi)\times
\]
\be
\times d\nu_1 d\nu_2 d\nu d\xi_1d\xi_2d\xi,  \n{4.25}
\ee
\[
P(B;C)=\left(\gD x
\right)^2 \int|\xi_1\xi_2|P(\nu_1,\eta_1=0,\xi_1, \nu_2,\eta_2=0,\xi_2) d\nu_1
d\nu_2 d\xi_1d\xi_2.
\]
The limits of the integration in Eq.(4.25) are\\
1) for $\nu_1$, $\nu_2$, $\nu$ -- from $\nu_t$ to $+\infty$,\\
2) for $\xi_1$, $\xi_2$ -- from $-\infty$ to $0$,\\
3) for $\xi$ -- from $0$ to $+\infty$.\\
The conditional probability is the ratio of these two integrals:
\be
\frac{P(A;B;C)}{P(B;C)}=N^+_{min}(x)\gD x,  \n{4.26}
\ee
where $N^+_{min}(x)\gD x$ is the differential density of minima above $\nu_t$
(here and below $x$ and $r$ are in the units of $R_*$). The constraint
that $\nu(x)$ is a minimum of any arbitrary value leads to identical
equations except that the integration $d\nu$ is over all $\nu$ from
$-\infty$ to $+\infty$.

The number of minima $n^+_{min}(r)$ and the total number of minima
$n_{min}(r)$ of arbitrary value between $x_1$ and $x_2$ can be found by
substituting $\gD x$ in (4.21) by $dx$ and integrating from
$x_1$ to $x_2$:
\[
n^+_{min}(r)=\int_{x_1}^{x_2}N^+_{min}(x)dx,
\]
\be
n_{min}(r)=\int_{x_1}^{x_2}N_{min}(x)dx, \n {4.27}
\ee
where $N_{min}(x)$ is the differential density of minima of an arbitrary
value at the point $x$. The probability that any given minimum (from
this collection of minima) has a value above $\nu_t$ is, therefore,
$N^+_{min}(r)/N_{min}$. The probability that two given maxima (at
points $x_1$ and $x_2$) are connected together is equal to the
probability that all $n_{min}(r)$ minima between them are above
$\nu_t$. All points of considered minima can be labeled from
$x_1^{min}$ and $x_{n(r)}^{min}$ and the desired probability is
\be
P(x_1\laar x_2)=\frac{N^+_{min}(x_1^{min})}{N_{min}(x_1^{min})}\hspace{0.2cm}
... \hspace{0.2cm} ... \hspace{0.2cm}
\frac{N^+_{min}(x_{n(r)}^{min})}{N_{min}(x_{n(r)}^{min})}\ . \n{4.28}
\ee
Here, $P(x_1\laar x_2)$ is the probability that  $\nu(x)>\nu_t$, for $x_1\le
x\le x_2$ (i.e. the probability that $x_1$ and $x_2$ are connected).
In order to estimate this value, we substitute $n_{min}^+(r)/r$ and
$n_{min}(r)/r$ by $N^+_{min}(x_i)$ and $N_{min}(x_i)$ respectively
and Eq.(4.28) simplifies to:
\be
P(x_1\laar x_2)\approx\left(\frac{n_{min}^+(r)}{n_{min}(r)}
\right)^{n_{min}(r)}\ . \n{4.29}
\ee
Note, that we always have $n_{min}(r)\ge 1$ and $n_{min}^+(r)/n_{min}(r)\le
1$.  If we consider two maxima which are spaced at $r\gg 1$, then
$P(A;B;C)=P(A)\cdot P(B)\cdot P(C)$, and $P(B;C)=P(B)\cdot P(C)$
because events $A$, $B$, $C$ become independent. In this case we have:
\be
\begin{array}{lc}
N^+_{min}(x)=\widetilde{N}_{min}^+={\normalsize{const}},\\
N_{min}(x)=\widetilde{N}_{min}={\normalsize{const}},\\
n^+_{min}(r)=N^+_{min}\cdot r,\\
n_{min}(r)=N_{min}\cdot r,
\end{array}
\n{4.30}
\ee
where $\widetilde{N}^+_{min}$ and   $\widetilde{N}_{min}$ are the
number of minima above $\nu_t$ and the number of minima of arbitrary
height respectively, without the condition that there are maxima at
$x_1$ and $x_2$. Combining Eqs.(4.29) and (4.30) we get
\be
P(x_1\laar x_2) = \left(\frac{N^+_{min}}{N_{min}}\right)^{N_{min}\cdot r}\ .
\n{4.31}
\ee
If $r\le 1$, then most probably only one minimum is located between two
maxima $(n_{min}(r)=1)$ and
\be
P(x_1\laar  x_2)\approx N_{min}^+(r/2)\ . \n{4.32}
\ee

\begin{center}
{\bf c. Shape around two maxima}
\end{center}
The expected value of the field around two maxima at the point $x_1$
and $x_2$ can be obtained from the covariance matrix of
$7\times7$ variables: $\nu_{1,2}=\nu(x_{1,2})$;
$\eta_{1,2}=\eta(x_{1,2})$; $\xi_{1,2}=\xi(x_{1,2})$ and $\nu(x)$. The
derivation is given in detail in Appendix A2. The steps are as follows:\\
1) The conditional probability that the field in point $x$ falls in
the range from $\nu$ to $\nu+d\nu$ on condition that
$\nu(x_{1,2})=\nu_{1,2}$, $\eta(x_{1,2})=0$, $\xi(x_{1,2})=\xi_{1,2}$,
is
\be
P(\nu|B;C)d\nu=\frac{P(\nu(x_{1,2})=\nu_{1,2},\hspace{0.2cm} \eta(x_{1,2})=0,
\hspace{0.2cm}\xi(x_{1,2})=\xi_{1,2},\hspace{0.2cm} \nu)}
{P(\nu(x_{1,2})=\nu_{1,2},\hspace{0.2cm} \eta(x_{1,2})=0,\hspace{0.2cm}
\xi(x_{1,2})=\xi_{1,2})}d\nu, \n{4.33}
\ee
where events $B$ and $C$ are the conditions that there are maxima at
the points $x_1$ and $x_2$ respectively.\\
2) Multiplication of the right hand side of Eq.(4.33) by $\nu$  and
integration over $\nu$ gives the expected value
of the field at an arbitrary point $x$, $\lan\nu(x)\ran$. Multiplication of
the right hand side of Eq.(4.33) by $\nu^2$ and integration over $\nu$
results in the variance $\lan\nu(x)^2\ran$.\\
The result (shown in Fig.4) can be applied to the local
analysis of one-dimensional data, since we can estimate the
probability for an appropriate pair of extrema to be Gaussian or not.

\begin{center}
{\bf d. Two-point peak-peak correlation function}
\end{center}

If a maximum with the height above $\nu_t$ is located at the point
$r=0$, then the differential density of maxima $N_{max}^+(r)$ above
this threshold at the distance $r$ from the given maximum can be
calculated using the technique of subsection {\bf b}:
\be
N^+_{max}(r)dr=\frac{\int|\xi_1\xi|P(\nu_1,\eta_1=0,\xi_1, \nu,\eta=0,\xi)
d\nu_1 d\nu d\xi_1d\xi}{\int|\xi_1|P(\nu_1,\eta_1=0,\xi_1) d\nu_1d\xi_1}
\frac{\gs_2}{\gs_1}dr\ . \n{4.34}
\ee
This density has to be compared with  $\widetilde{N}_{max}^+$, i.e.
with the  density of maxima above $\nu_t$ without the condition that
there is a maximum at $r=0$:
\be
\widetilde{N}^+_{max}=\int|\xi|P(\nu,\eta=0,\xi)d\nu d\xi. \n{4.35}
\ee
Then, the two-point peak-peak correlation function is:
\be
\Psi_{p-p}(r)=
\frac{N^+_{max}(r)-\widetilde{N^+}_{max}}{\widetilde{N^+}_{max}}. \n{4.36}
\ee
The two-point peak-antipeak correlation function can be obtained
similarly to Eq.(4.36):
\be
\Psi_{p-ap}(r)=
\frac{N^+_{min}(r)-\widetilde{N^+}_{min}}{\widetilde{N^+}_{min}}\ .\n{4.37}
\ee

\begin{center}
{\bf e. Clusters of peaks}
\end{center}

The problem of clusterisation of peaks  has a few difficult
points.  Calculation of the probability of appearance of a cluster of
length $k$ is quite difficult when
peak-peak and peak-antipeak correlations are taking into account.
However the calculation of the mean length of a cluster for an
appropriate level $\nu_t$ is much easier and can be done analytically.

Let us consider a maximum above the threshold $\nu_t$ at the point
$x=x_*$, and calculate the total numbers of maxima and minima above
$\nu_t$ in the vicinity of this peak for $x$ from $x_*-r$ to $x_*+r$:
\be
n^+_{max}(r)=\widetilde{N}_{max}^+\int_{x_*-r}^{x_*+r}(\Psi_{p-p}(x)+1)dx,
\n{4.38}
\ee
\[
n^+_{min}(r)=\widetilde{N}_{min}^+\int_{x_*-r}^{x_*+r}(\Psi_{p-ap}(x)+1)dx.
\]
In the vicinity of this maximum clusters of the different length $k$
appear with the appropriate probability, which is difficult to determine.
However, the total number of clusters in this vicinity is given exactly by
\be
\sum_kN_k(r)=n^+_{max}(r)-n^+_{min}(r)+1\ , \n{4.39}
\ee
where $N_k$ is the number of clusters of the length $k$. Obviously,
the summation of the number of maxima in all clusters gives the total
number of maxima above $\nu_t$  from $x_*-r$ to $x_*+r$:
\be
\sum_k kN_k=n^+_{max}(r)+1. \n{4.40}
\ee
The unit in the right hand side of Eqs.(4.39) and (4.40)
accounts for one additional maximum at $x=x_*$. Therefore, the mean
length of a cluster $\ov{k}(r)$, is:
\be
\ov{k}(r)=\frac{\sum_k k\cdot N_k(r)}{\sum_k N_k(r)}=
\frac{n^+_{max}(r)+1}{n^+_{max}(r)-n^+_{min}(r)+1}. \n{4.41}
\ee
Note, that if $n^+_{max}(r)=n^+_{min}(r)$ (with $n^+_{max}(r)\ge
n^+_{min}(r)$ always being true), then only one cluster of the length
$n^+_{max}(r)+1$ appears. We are interested in the mean value $\ov{k}$
for the whole considered region $[0;L]$. This value can be found by
using Eq.(4.38) and  Eq.(4.41) and averaging over all
maxima, and letting the integration in  Eq.(4.38) go from 0 to $L$. If
the statistical properties of the region are ``good enough'' (namely
$r_c\ll L$), then
\be
n^+_{max}(r)=\widetilde{N}_{max}^+\int_0^L(\Psi_{p-p}(x)+1)dx \approx
\widetilde{N}_{max}^+L\gg 1, \n{4.42}
\ee
\[
n^+_{min}(r)=\widetilde{N}_{min}^+\int_o^L(\Psi_{p-ap}(x)+1)dx \approx
\widetilde{N}_{min}^+L\gg 1,
\]
and
\be
\ov{k}=
\frac{\widetilde{N}_{max}^+}{\widetilde{N}_{max}^+-\widetilde{N}_{min}^+}=
\frac{1}{1-\ga(\nu_t,\gc)}\ , \n{4.43}
\ee
where $\ga(\nu_t,\gc)=\widetilde{N}_{min}^+/\widetilde{N}_{max}^+$.
The values $\widetilde{N}_{max}^+(\nu_t)$ and
$\widetilde{N}_{min}^+(\nu_t)$ can be found analytically, see Rice (1944,
1945):
\be
\widetilde{N}_{max}^+=\frac{1}{\sqrt{(2\pi)^3(1-\gamma^2)}}
\frac{\gs_2}{\gs_1} \int_{\nu_t}^{\infty}d\nu\int_{-\infty}^0|\xi|
e^{-\frac{\nu^2}{2}-\frac{(\xi+\gamma\nu)^2}{2(1-\gamma^2)}}d\xi, \n{4.44}
\ee
\be
\widetilde{N}_{min}^+=\frac{1}{\sqrt{(2\pi)^3(1-\gamma^2)}}
\frac{\gs_2}{\gs_1}\int_{\nu_t}^{\infty}d\nu\int_0^{\infty}|\xi|
e^{-\frac{\nu^2}{2}-\frac{(\xi+\gamma\nu)^2}{2(1-\gamma^2)}}d\xi. \n{4.45}
\ee
The result of this integration is presented in Appendix A1. The
integration over $d\xi$ in Eq.(4.44) gives the differential density of
maxima and minima which are plotted in Fig.5a for different
values of $\gc$.  From Eq.(4.43) we can see that the mean length of the
clusters depends only on the density of minima and maxima above
the threshold $\nu_t$. For a large value of $\nu_t$, $\ov{k}\ra 1$
because, as we see from Fig.5a, almost all the high extrema are
maxima and $\ga\ll1$. For a large negative value of $\nu_t$
$\ov{k}\ra\infty$, because, in this case,
$\widetilde{N}_{max}^+(\nu_t)\ra\widetilde{N}_{min}^+(\nu_t)$ and
$\ga\ra 1$.

The dependence of $\ov{k}$ on the level of slice $\nu_t$   for
different $\gamma$ are plotted in Fig.5b. A low value of $\gamma$
corresponds to the high rate of clusterisation. Note, that $\gamma$
lies in the interval $0<\gamma\le1$. The value $\gc=1$ corresponds to
the power spectrum $P(k)={\normalsize const}\cdot\gd(k-k_0)$.  In this
case we have $\ov{k}=1$ for  $-|\nu_0|<\nu_t<|\nu_0|$; $\ov{k}=\infty$ for
$\nu_t<-|\nu_0|$; $\ov{k}=0$ for $\nu_t\ge|\nu_0|$. Here, $\nu_0$ is the
random Gaussian value and depends on the appropriate realization.

\subsection{Two-dimensional field}

In this section, we develop the theory of the statistical properties of the
two-dimensional field. These properties can be applied to further analysis
of the sky maps. In what follows we obtain an important characteristic
statistical
feature of the two-dimensional Gaussian field: clusterisation of peaks
depends only on one spectral parameter $\gc$, just as in the one-dimensional
case. This spectral parameter of a Gaussian field of CMB radiation
depends effectively only on the model of recombination. Note,
that this natural feature of the Gaussian field can be used not only
for determination of the ionization history, but also as a test of the
presence of noise in a map (which, most probably, is non-Gaussian).
We derive equations for the field in the vicinity of two neighbouring maxima.
These equations define the details of the structure of the cluster in
the vicinity of the maxima and can be used for the local analysis of the
appropriate part of the map.
\begin{center}
{\bf a. Surface $\gD T(x,y)$}
\end{center}

A two-dimensional field can be described as a two-dimensional surface in
a three-dimensional space, $\gD T(x,y)$, Fig.6. We will cut this surface at
the different levels $\nu_t$. The surface intersects the plane $\gD
T(x,y)=\nu_t$ along the lines of level $\nu_t$ (see Map.2a,b). If, for
example, two local maxima with the values above this threshold $\nu_t$
are confined by a closed line of level $\nu_t$, then these maxima are
considered  connected together in one cluster. It is obvious that two
neighbouring maxima on this surface can be connected together only
through the saddle point between them, see Map.2. Therefore, if the saddle
point
between two maxima is above the threshold $\nu_t$, then these maxima are
connected together. Let us define a cluster of the length $k$ similar to
the one-dimensional case.

{\bf Definition:} Cluster of the length $k$ is a collection of $k$
maxima confined by a closed  level-line.

In what follows we describe the clusterisation of maxima, namely, we
find the mean length of the clusters as a function of the level  of the
slice, $\nu_t$, and spectral parameter, $\gamma$.

\begin{center}
{\bf b. Extremal points}
\end{center}

Let us introduce the following expressions:
\be
\begin{array}{lc}
\nu=\frac{\gD T}{\gs_0};\hspace{0.5cm} \eta_1=\frac{\gD
T'_x}{\gs_1};\hspace{0.5cm}\eta_2=\frac{\gD T'_y}{\gs_1};\\
\xi_{11}=\frac{\gD T''_{xx}}{\gs_2};\hspace{0.5cm} \xi_{22}=\frac{\gD
T''_{yy}}{\gs_2};\hspace{0.5cm} \xi_{12}=\frac{\gD T''_{xy}}{\gs_2}.
\end{array}
\n{4.46}
\ee

If  $\eta_1=\eta_2=0$ at the point $(x_0,y_0)$, then this point is an
extremum, and the field $\nu$ in the vicinity of $(x_0,y_0)$ is
\be
\nu(x,y)=\nu(x_0,y_0)+\xi_{11}x^2+2\xi_{12}xy+\xi_{22}y^2. \n{4.47}
\ee
We rotate the $x$, $y$ axes by the angle $\gff=\ha
{\normalsize{arctg}}\frac{2\xi_{12}}{\xi_{11}-\xi_{22}}$ around the point
($x_0, y_0$)
and find the eigen values of the second derivatives matrix $\xi$,
$\gl_1$ and $\gl_2$:
\be
\begin{array}{lc}
\xi_{11}=\gl_1\cos^2\gff+\gl_2\sin^2\gff,\\
\xi_{22}=\gl_1\sin^2\gff+\gl_2\cos^2\gff,\\
\xi_{12}=(\gl_1-\gl_2)\sin\gff\cos\gff.\\
\end{array}
\n{4.48}
\ee
All extrema can be divided in three types: maxima $\gl_{1,2}<0$, saddle
points ($\gl_1>0,\gl_2<0$) or ($\gl_1<0,\gl_2>0$), and minima
$\gl_{1,2}>0$.

\begin{center}
{\bf c. Density of maxima, minima and saddle points.}
\end{center}

Derivation of the density of saddle points is given in Appendix B1. (The
density of maxima and minima for the
two-dimensional Gaussian field was obtained by B.E., 1987). Here
we present the results of our analytical calculations for the saddle
points and mention the results of calculations by B.E. for the maxima and
minima. The
differential density of the saddle points is:
\be
N_{sad}(\nu,\gamma)d\nu=\frac{1}{\sqrt{32\pi^3}}\frac{\gs_2^2}{\gs_1^2}
\frac{1}{\sqrt{3-2\gamma^2}} e^{-\frac{3\nu^2}{2(3-2\gamma^2)}}d\nu\ . \n{4.49}
\ee
The integration of Eq.(4.49) from $\nu_t$ to infinity gives the number of
saddle points above the same threshold $\nu_t$
\be
n_{sad}(\nu_t)=\int^{\infty}_{\nu_t}N_{sad}(\nu)d\nu=\frac{1}{8\pi\sqrt{3}}
\frac{\gs_2^2}{\gs_1^2}\left[1-\gF\left(\frac{\nu_t\sqrt{3}}
{\sqrt{2(3-2\gc^2)}}\right)\right]\ . \n{4.50}
\ee
where $\gF (x)$ is the probability integral by Gradshteyn and Ryzhik, 1980.
The full density of saddle points is:
\be
n_{sad}(-\infty, \gamma)=\frac{1}{4\pi\sqrt{3}}\frac{\gs_2^2}{\gs_1^2}.
\n{4.51}
\ee
The differential density of maxima is:
\[
N_{max}(\nu)d\nu=\frac{\sqrt{2}}{4\pi\sqrt{\pi}}
\frac{\gs_2^2}{\gs_1^2}\cdot
\left\{\frac{1}{4\sqrt{3-2\gc^2}} e^{-\frac{3\nu^2}{2(3-2\gc^2)}}
\left[1+\gF\left(\frac{\gc\nu}{\sqrt{(2-2\gc^2)(3-2\gc^2)}}\right)\right]
\right.
\]
\be
\left.+\frac{\gc(\nu^2-1)}{4}e^{-\frac{\nu^2}{2}}
\left[1+\gF\left(\frac{\gc\nu}{\sqrt{2(1-\gc^2)}}\right)\right]
+\frac{\gc\nu\sqrt{1-\gc^2}}{\sqrt{2\pi}}e^{-\frac{\nu^2}{2(1-\gc^2)}}\right\}\
. \n{4.52}
\ee
Note, that $N_{min}(\nu)=N_{max}(-\nu)$ and
\be
n_{max}(\nu_t,\gc)=\int_{\nu_t}^{\infty}N_{max}(\nu,\gc)d\nu
\hspace{1cm} n_{max}(-\infty,\gc)=
\frac{1}{8\pi\sqrt{3}}\frac{\gs_2^2}{\gs_1^2}\ . \n{4.53}
\ee
The distribution of maxima, minima and saddle points are shown in
Fig.7a for different values of $\nu$. As it can be seen from this
figure,  this distribution is very sensitive to the value of $\gc$. In
what follows, we demonstrate the importance of sensitivity of
$N_{sad}(\nu,\gc)$, $N_{max}(\nu,\gc)$ and  $N_{min}(\nu,\gc)$ to the value
of $\gc$.  Namely, the rate of clusterisation of peaks in a random
two-dimensional Gaussian field depends only on the value of $\gc$. Note
that the corresponding  behaviour of  $N_{max}(\nu,\gc)$ and
$N_{min}(\nu,\gc)$ was discussed earlier in Section 3.1 for the
one-dimensional cross-section.  In the one-dimensional case, the rate of
clusterisation (in a random Gaussian field) also depends only on the
parameter $\gc$ as we demonstrated.

\begin{center}
{\bf d. Clusters of peaks}
\end{center}

Let us consider an appropriate realization of the two-dimensional
random Gaussian process in a map $l\times l$, see Map.2. In this
figure the maxima and the saddle points with values larger than the
level of the slice, $\nu_t$, together with the isolines $\nu=\nu_t$ are shown.
If
the value of $\nu_t$ is high, all maxima are separated and only
clusters of the length $k=1$ are observed.  Reduction of the level
$\nu_t$  leads to the appearance of big clusters (maxima begin to connect
together and generate clusters). Let us consider a slice of the map and
calculate the number of clusters $N_k$ of the length $k$ for
$k=1,2,...$.  Summation of the number of maxima  in all clusters of the
map gives the total number of maxima:
\be
\sum_{k=1}^{\infty}N_k\cdot k=n_{max}(\nu_t). \n{4.54}
\ee
Now, let us calculate the total number of clusters (having arbitrary
number of maxima). Each cluster can contain maxima, minima and saddle
points. The numbers of maxima, minima and saddle points in one cluster
are not independent values.  If one particular cluster contains
$k=k_{max}$ maxima and $k_{min}$ minima, then it has to contain
$k_{sad}=k_{max} + k_{min}-1$ saddle points. Therefore, the total number of
clusters of arbitrary length $k$ for a given slice - level $\nu_t$ is:
\be
\sum_k N_k=n_{max}(\nu_t) + n_{min}(\nu_t) -n_{sad}(\nu_t)\ . \n{4.55}
\ee
Using Eqs.(4.54) and (\ref{4.55}), one can write:
\be
\ov{k}=\frac{\sum kN_k}{\sum N_k}=\frac{n_{max}}{n_{max}+n_{min}-n_{sad}}\ .
\n{4.56}
\ee
If only high levels of $\nu_t$ are considered, then the number of minima,
$n_{min}(\nu_t)$, is small for $\gc$ sufficiently different from zero,
(see Fig.7a) and can be disregarded.

In this case, two neighbouring maxima are counted in one cluster for the
level $\nu_t$, only if the saddle point between them is above this
level. Therefore, each cluster  containing $k$ maxima, also contains
$k-1$ saddle points (clusters of the length 1 does not contain any
saddle point). In this case the mean length of the clusters is
\be
\ov{k}\approx 1 +\ga(\nu_t,\gc), \n{4.57}
\ee
where
\be
\ga(\nu_t,\gc)=\frac{n_{sad}(\nu_t,\gc)}{n_{max}(\nu_t,\gc)}, \n{4.58}
\ee
similar to the one-dimensional case. The ratio $n_{sad}/n_{max}$ depends
not only on the level of $\nu_t$, but also on spectral parameter
$\gc$.  Thus, $\ga=\ga(\nu_t,\gc)$ also depends on both $\nu_t$ and
$\gc$. The dependence of $\ov{k}$ on the level $\nu_t$ in the general case
Eq.(4.57) is shown in Fig.7b, for different values of $\gc$. For
low values of $\gc$ the clusterisation occurs faster than for high
values of $\gc$. This can be concluded from Fig.7a, where the
distributions of extrema for different values of $\gc$ are shown.
Therefore, clusterisation of peaks in a two-dimensional random Gaussian
field (namely the mean length of clusters) can be described in
terms of only one parameter, $\gc$.

For a high level of the slice, almost all extrema are maxima, and
$n_{min}\ra 0$ and $n_{sad}\ra 0$ ($\ov{k}\ra 1$ and we have separated maxima).
For $\nu_t\ra 0$, $n_{max}+n_{min}-n_{sad}$ tends to zero (see Appendix
B1), and $\ov{k}\ra \infty$. Infinite $\ov{k}$ is equivalent to the
existence of an infinite cluster of maxima. However, the  finite  size
of the region considered prevents the realization of an infinite
cluster (a finite region has only a finite number of maxima). In an
infinite region, the existence of an infinite cluster and the
percolation are essentially equivalent. Since percolation is also
defined for a finite region, the concept of an infinite cluster in a
finite region can be addressed in terms of percolation.

\begin{center}
{\bf e. Percolation}
\end{center}

Let us colour the map of the considered region in two colours: black and
white. The black area is for $\nu>\nu_t$, and the white area is for
$\nu<\nu_t$ (see Map.3). Percolation through black (white) zones is
usually defined as the possibility to ``walk'' from one side of the map
to another (from the lower to the upper border and from the left to
the right side) only through the black (white) zones. Roughly speaking,
percolation means the existence of a black (white) cluster with the
linear size of the considered map, or the existence of an infinite
cluster in the case of an infinite map. In terms of extremal points,
percolation over black (white) zones means the existence of an infinite
cluster of maxima (minima).  The change of percolation over black zones
into percolation over white zones has to take place at the $\nu_p=0$ in
an infinite map. Gaussian nature of the distribution of the CMB
anisotropy assures this property.  In a particular realization of the
Gaussian process in a finite map a change of percolation regimes arises
when $\nu$ is slightly different from zero, because of the statistical
character of the process. The difference of $\nu$ from zero can be
evaluates as follows:\\
1. Each hotspot (black zone)
and coldspot (white zone) can be considered as an approximately independent
realization (correlations between regions of the size $r_c\times r_c$ is
negligible).\\
2. The total number of such approximately independent realizations is
$N\sim L^2/r_c^2.$\\
3. The variance is $D\sim 1/\sqrt{N}\sim r_c/L$.

Therefore,  for the appropriate region of the sky the
probability of $\gD T$ fluctuations to be Gaussian can be estimated. This can
be accomplished using  the percolation technique: if $\nu_p\gg r_c/L$ then,
most probably, the additional non-Gaussian noise is present
in this map.

Note, that this obvious test can be considered only as the first step of
the detection of the noise, because the critical point of percolation
$\nu_p$  can be equal to zero even with non-Gaussian noise (for
example, if the noise is symmetrical relatively to $\nu=0$).

\begin{center}
{\bf f. Shape around two maxima}
\end{center}

The expected value of the field around two given maxima in the two-dimensional
case can be derived similarly to one-dimensional case
(Section 4.1). The derivation is given in Appendix B2. This result
describes the clusterisation of two neighbouring maxima and can be
applied to the local analysis of  small patches of the map, which
contain only the area of two neighbouring hotspots.

This area can be checked for presence of noise  by evaluating the
probability  that these peaks  have a Gaussian nature, as it  is
explained in the following.  As it was mentioned (e.g. by B.E.) the
contour curves in the neighbourhood of Gaussian peaks are ellipses.
The directions of major and minor axes are different for different
maxima.  Thus, two neighbouring maxima have some relative orientation.
Suppose that, in an appropriate part of the map, these maxima are
located at the distance $r$ from each other, and  that the value of the
field and its second derivatives (relative orientations) are $\nu_1$,
$\xi^1_{11}$,  $\xi^1_{12}$, $\xi^1_{22}$ and $\nu_2$, $\xi^2_{11}$,
$\xi^2_{12}$, $\xi^2_{22}$ for the first and the second maximum
respectively. Then it is possible to predict the mean value of the
field and its variance in the vicinity of given maxima and to compare
this result to observed pairs of hotspots in the map.

The mean value of the field around two maxima is presented in
Fig.8, for
different kinds of spectra (i.e., for the different values of the spectral
parameter $\gc$). Note, that this value depends on the behaviour of the
correlation function on the scale of this area $\ge r_c$ mainly being
determined by
$\gc$. For low values of $\gc$ two neighbouring hotspots stick
together faster than for high values of $\gc$. This fact is in good
agreement with the results of subsection {\bf d}. The increase of the rate
of clusterisation for decreasing values of $\gc$ is favorable to
faster production of big clusters.

\section{How  can we analyse the sky maps?}

In this section we propose a method of analyzing of the observational
data on CMB anisotropy for one- and two-dimensional experiments.
Interpretation of observations of a specific patch of sky
is complicated for the following reasons:\\
1. Most of the cosmological models predict the CMB anisotropy to be a
single realization of a random Gaussian process for which the properties can be
predicted only for an ensemble average of such realizations. Any single
realization differs from its ensemble average approximately by the
value of the dispersion (the so-called ``cosmic variance'').\\
2. Presence of a non-Gaussian noise can be interpreted as fluctuations
of CMB itself and can give a wrong contribution to the correlation
function or, equivalently, to the power spectrum.

As for the first reason, the following comments can be made. A  small
part of the sky may not be representative. This may lead to
misinterpretation  of properties of the two-point
correlation function of the CMB as obtained from one patch only.  Thus, the
finite nature of the considered region may lead to misinterpretation of
the correlation function of e.g. the galaxy
distribution on scales  comparable to the size of the region.
This effect causes difficulties in detection of the so-called
``long distance correlation'' in the galaxy distribution which can
appear in cosmological models with non-standard inflation (Starobinsky, 1992).

As for the second reason, the correlation function obtained by
measuring the sum of the relic radiation itself and non-Gaussian noise
will of course be the correlation function of the sum of the noise
and relic radiation.

Note, that both of these two reasons may lead to a
correlation function which closely can imitate the presence of a
Doppler-peak in the spectrum of perturbations of CMB. In this chapter,
we propose the topological method of analysis of the sky maps
which is useful for (1) detection of non-Gaussian noise,(2)
detection of a Doppler-peak in the spectrum.

\begin{center}
{\bf a. One dimensional experiment}
\end{center}

Assume that we have data from a one-dimensional experiment being a
one-dimensional slice of the two-dimensional realization
(Fig.4, Section 4.1). If the primordial signal is the result of
perturbations of the CMB radiation only, then the correlation analysis
is equivalent to the topological analysis because all statistical
properties of the Gaussian field are completely characterized by its
power spectrum. In this case, we should only estimate the dispersion of
the correlation function (section 2) on the most interesting scales
(about $15'$ to $ 1^o$), which is determined by the size of the
considered region (Hinshaw et al., 1995). This is necessary for the
estimation of the probability that the peak-like anomaly (if it is
detected in the correlation function of the observed data) occurs due
to the presence of the Doppler-peak in the spectrum, and not due to
``bad'' statistical properties of the selected region.

Next, the rate of the clusterisation of peaks (Section 4.1) for
different values of the filter (Section 3) can be checked. It is
obvious that the clusterisation of minima is equivalent to the
clusterisation of maxima if the sign of fluctuations is changed.
This equivalence exists
due to the symmetrical statistical properties of the Gaussian field
with respect to zero. Variation of the value of the filter leads to a
change of the spectral parameter $\gc$. For a given parameter $\gc$,
the rate of clusterisation of peaks is determined only by the Gaussian
nature of the field; this rate is plotted on (Fig.5b), for different values
of $\gc$. Note, that for different cosmological models, the dependence on
$\gc(r_0)$ (where $r_0$ is the resolution) is different (Fig.3a,b).
If, for each value of $\gc$ (which corresponds to an appropriate value
of $r_0$), the rate of clusterisation corresponds to the Gaussian
statistic (Fig.5b), then this indicates a Gaussian random field and the
observed
correlation function is the correlation function of CMB.

Now let us discuss how the situation will change in case of
non-Gaussian noise being present. In this case, the correlation and
the topological analysis are not equivalent to each other. The presence
of the noise shows up as presence of ''noise peaks`` in the
random field. These ``noise peaks'' give a contribution to the observed
two-point correlation function. If the observed correlation function
is not analysed properly, the presence of the ``noise peaks'' in the
observed data may lead to wrong conclusions. In order to avoid a
wrong interpretation of the  observed data, we propose to perform the
cluster analysis:\\
1) The rate of clusterisation (as discussed
above) should be checked. If this test shows us that the rate does not
correspond to one spectral parameter $\gc$, then additional random
noise is present in the data.\\
2) A search for ``noise peaks'' should be performed.\\
If these tests indicate the presence of noise, then how can
we remove the ``noise peaks''? We propose to check the probability
to be Gaussian for each pair of neighbouring extrema, by using the
technique of Section 4.2, Fig.4. The peaks with small probability
should not be considered in the further analysis.

\begin{center}
{\bf b. Two-dimensional data}
\end{center}

Difficulties in the interpretation of the two-dimensional data are
similar to that in one-dimensional case, but the statistical properties
are usually  much better. The analysis can be analogous to that
for the one-dimensional case, and the steps are as follows:\\
1) Filtering of the sky map, see Section 3. This step can be useful
for the following reasons. The use of the specific (for example
non-Gaussian) filtering removes the influence of long modes and,
therefore, leads to a decrease of the correlation radius. Since we
are interested only in the relatively high modes, such a filtering
improves the statistical properties of the map (for the purpose of the
cluster analysis).  A filtered map contains a larger number of peaks and is
more representative than the unfiltered one. Thus we perform the
cluster analysis on the filtered map.\\
2) Cluster analysis. This step consists of three substeps:\\
a. Percolation as the first test of the presence of non-Gaussian
noise (see Section 4.2, Map.3).\\
b. Rate of clusterisation. This substep can be performed similarly
to the one-dimensional case, but using the results of Section
4.2, Fig.7b.\\
c. Detection of the ''noise peaks'' (if the previous test indicates the
presence of noise), as for the one-dimensional case, but using the
results of Section 4.2, Fig.8.

\begin{center}
{\bf c. Topology for different models }
\end{center}
As it was shown above, the most important topological properties
(namely clusterisation of peaks in a random Gaussian field) can be
explained in terms of only  one spectral parameter $\gc$.  This spectral
parameter is very sensitive to the presence of Doppler-peaks in the
spectrum of fluctuations of $\gD T$ for the resonance filter
(Section 3). It means that the rate of clusterisation will change from
one model to the next, for the same value of the filter. A big value of
$\gc$ corresponds to a low rate of clusterisation (Fig.7b) and, for
the same level of the slice, the expected number of big clusters is
less than for a small value of $\gc$. For the model with a
pronounced Doppler-peak, the value of $\gc$ is larger than it is for the
model without a Doppler-peak in the spectrum (Fig.3a,b). Therefore, for such
models, the clusterisation happens more slowly than for the model
without a Doppler-peak (Map.4a,b). Thus, the number of big clusters for
the model without a Doppler-peak is larger than for the model with a
Doppler-peak.

Another obvious difference consists in that a number of maxima
(minima) per unit square is greater for the model with a high
Doppler-peak, because the presence of this peak leads to a decrease of
the correlation angle.

\section{Discussion}

The results of this paper can naturally be separated into two parts.
One is the  presentation of
new theoretical results on the theory of clusterisation of peaks in
a Gaussian random field and on the theory of percolation. All
mathematical results are analytical. The other is to suggest how such
analytical calculations can be applied to the analysis of the observational
data of the CMB anisotropy.

\subsection{Theory} The most important new results on the
clusterizarion of peaks in a Gaussian random filed, in this paper, are the
following:

\begin{center}
{\bf a. One-dimensional cross-section of the
two-dimensional random field}
\end{center}
\noi 1. Analytical
calculations of the mean length of the cluster of peaks in,
one-dimension, Eqs.(4.43-4.45).\\
2. The shape of the random field in the vicinity of two neighbouring
maxima, namely the mean value of the field $\lan\nu(x)\ran$ and its
dispersion $\lan\nu(x)\ran\pm\sqrt{\lan\nu^2(x)\ran}$  about this mean.
These values are to be determined from the Gaussian probability
distribution of the value of $\nu(x)$, given that there are two maxima
at the points $x_1$ and $x_2$ of given heights and curvatures (second
derivatives), Eqs.(4.33, A2.8, A2.9).

\begin{center}
{\bf b. Two-dimensional  field}
\end{center}
\noi
1. Calculation of the densities of saddle points: differential in height
$N_{sad}(\nu)d\nu$ ,Eq.(4.49), and integral (above the threshold $\nu_t$),
Eq.(4.50).\\
2. The shape around two neighbouring maxima similar to the one-dimensional
case,
that is the expected average value and its dispersion, (Eqs.(B2.5,B2.6)).\\
3. Calculation of the mean length of the cluster of peaks in the
two-dimensional case, Eq.(4.56).\\
4. We find, that the level of percolation, $\nu_p$, and the level
of appearance of an infinite cluster of peaks are the same:
$\ov{k}(\nu_p)=\infty$. It is well known, that in the two-dimensional case
we have $\nu_p=0$ (because of the symmetrical properties with
respect to the zero). Similar analytical results can be obtained for the
3-dimensional case (now it is known only numerically), but
is most probably only of theoretical interest and is beyond
the scope of this paper.

\subsection{Applications to the analysis of the observational data.} In
Section 5 we have presented our  suggestions how analytical calculations
can be used in the analysis of observational data. The proposed
topological method can provide more information about the observational
data than the correlation analysis.

\section{Acknowledgements} The authors wish to thank P.Coles, B.Jones,
P.Naselsky, I.Novikov, J.Ostriker, M.Rees for the helpfull discussions
and comments, and E.Kotok for help in computations and preparation of
the paper. We are gratefull to TAC for the arrangment of an excellent
Cosmological workshop (May, 1995), in the framework of which the main
part of this work was done. Furthermore, D.N. is grateful to the staff of TAC,
Copenhagen Astronomical Observatory, and NORDITA for providing
excellent working conditions. This investigation was supported in part by
the Russian Foundation for Fundamental Research (Code 93-02-2929) and by a
grant ISF MEZ 300 as well as by the Danish Natural Science Research Council
through grant No. 9401635 and also in
part by Danmarks Grundforskningsfond through its support for the
establishment of the Theoretical Astrophysics Center.

\section{Appendix A1}

{\bf Density of maxima and minima in one-dimension}

The problem of the density of extrema is especially easy in one-dimension,
and was solved by Rice (1944, 1945). For a review see also
J.Bardeen, J.Bond, N.Kaiser and A.Szalay (1986, BBKS).

The differential density of maxima and minima is
\[
N_{max}(\nu)d\nu=
\]
\[
\frac{1}{(2\pi)^{3/2}}\frac{\gs_2}{\gs_1} \left\{\sqrt{1-\gc^2}
e^{-\frac{\nu^2}{2(1-\gc^2)}}+\gc\nu\sqrt{\frac{\pi}{2}}
e^{-\frac{\nu^2}{2}}\left[1+\gF\left(\frac{\gc\nu}{\sqrt{2(1-\gc^2)}}
\right)\right]\right\}d\nu, {\normalsize\hspace{0.1cm} (A1.1)}
\]
where $\gF(x)=\frac{2}{\sqrt{\pi}}\int_0^xe^{-t^2}dt$ is the probability
integral (Gradshteyn  and Ryzhik, 1980).
\[
N_{min}(\nu)=N_{max}(-\nu).
\]
The integrated density is:
\[
\widetilde{N}^+_{max}(\nu_t)=\int_{\nu_t}^{+\infty}N_{max}(\nu(d\nu)=
\]
\[
=\frac{1}{4\pi}\frac{\gs_2}{\gs_1}
\left\{1-\gF\left(\frac{\nu_t}{\sqrt{2(1-\gc^2)}}\right)+\gc
e^{-\frac{\nu_t^2}{2}}\left[1+\gF\left(\frac{\nu_t}{\sqrt{2(1-\gc^2)}}
\right)\right]\right\}. {\normalsize\hspace{1.1cm} (A1.2)}
\]
The density of maxima of arbitrary height is:
\[
\widetilde{N}^+_{max}(-\infty)=\frac{1}{2\pi}\frac{\gs_2}{\gs_1}.
{\normalsize\hspace{9.3cm} (A1.3)}
\]
\section{Appendix A2}

{\bf Shape around two maxima in one-dimension.}

Let us define the event $A$ as the presence of two maxima at the points
$x_1$ and $x_2$ respectively with parameters: $\nu(x_{1,2})=\nu_{1,2}$;
$\eta(x_{1,2})=0$; $\xi(x_{1,2})=\xi_{1,2}$, and the event $\nu$
with the field at the point $x$ located in the interval from
$\nu$ to $\nu+d\nu$. The joint probabilities are:
\[
P(\nu,B;C)d\nu(x)d\nu(x_1) ... d\xi(x_2)=P(\nu(x),\nu(x_1), ... ,
\xi(x_2)d\nu(x) d\nu(x_1) ... d\xi(x_2)\ ,
\]
\[
P(B;C)d\nu(x_1) ... d\xi(x_2)=P(\nu(x_1), ... , \xi(x_2)d\nu(x) d\nu(x_1)
... d\xi(x_2)\ ,
\]
\[
P(\nu,B;C)=\frac{1}{\sqrt{(2\pi)^7\det M_1}}e^{-Q_1},
{\normalsize\hspace{7.1cm} (A2.1)}
\]
\[
P(\nu,B;C)=\frac{1}{\sqrt{(2\pi)^6\det M_2}}e^{-Q_2},
\]
where $M_1$ and $M_2$ are the $7\times 7$ and $6\times 6$ covariance matrices,
respectively, $Q_1$ and $Q_2$ are the quadratic forms. The
conditional probability for $\nu(x)$ to be in the range from $\nu(x)$ to
$\nu(x)+d\nu(x)$ is:
$$
P(\nu|B;C)d\nu(x)=\frac{\int P(\nu,B;C)\gd(\nu(x_1)-\nu_1) ...
\gd(\xi(x_2)-\nu_2) d\nu(x_1)...d\xi(x_2)}{\int P(B;C)\gd(\nu(x_1)-\nu_1) ...
\gd(\xi(x_2)-\xi_2) d\nu(x_1)...d\xi(x_2)}d\nu(x).{\normalsize(A2.2)}
$$
Using (A2.1) and (A2.2) we obtain:
\[
P(\nu|B;C)d\nu(x)=\frac{1}{\sqrt{2\pi}}\sqrt{\frac{\det M_2}{\det M_1}}
e^{-(Q_1-Q_2)}d\nu(x).{\normalsize\hspace{4.4cm} (A2.3)}
\]
Now we define the functions:
\[
\Psi(r)=\frac{C(r)}{\gs^2_0};\hspace{1cm}\gff(r)=
\frac{C^{\prime\prime}(r)}{\gs^2_1};\hspace{1cm}
f(r)=\frac{C^{IV}(r)}{\gs^2_2};
\]
\[
\gd(r)=\frac{C^{\prime}(r)}{\gs_0\gs_1};\hspace{1cm}
\gl(r)=\frac{C^{\prime\prime\prime}(r)}{\gs_1\gs_2}\ .
{\normalsize\hspace{7.2cm}(A2.4)}
\]
In terms of these functions the covariance matrix $M_1$ is
\[
M_1=
\left(\begin{array}{ccccccc}
1 & \Psi & 0 & \gd & -\gamma/2 & \gamma\gff & \Psi_1\\
\Psi & 1 & -\gd & 0 & \gamma\gff & -\gamma/2 &  \Psi_2\\
0 & -\gd & 1/2 & -\gff & 0 & -\gl & -\gd_1\frac{x-x_1}{r_1}\\
\gd & 0 & 0 & 1/2 & \gl & 0 & -\gd_2\frac{x-x_2}{r_2}\\
-\gamma/2 & \gamma\gff & -\gff & \gl & 3/8 & f & \gamma\gff_1\\
\gamma\gff &-\gamma/2 & -\gl & 0 & f & 3/8 & \gamma\gff_2\\
\Psi_1 & \Psi_2 & -\gd_1\frac{x-x_1}{r} &  -\gd_2\frac{x-x_2}{r} &
\gamma\gff_1 & \gamma\gff_2 & 1
\end{array}
\right)\ ,
 {\normalsize\hspace{0.3cm} (A2.5)}
\]
where functions with subscripts ''1'', ''2'' and without subscripts are
evaluated at the points $r_1$, $r_2$ and $r$ respectively. Here
$r_1=|x-x_1|$, $r_2=|x-x_2|$, $r=|x_1-x_2|$.  $M_2$ can be obtained
from $M_1$ by excluding the last row and the last column. If we
introduce the new variables $y_i$, $i=1,6$, and $\widetilde{\nu}$, so
that
\[
y_1=\nu_1-\nu_2;\hspace{1cm} y_2=\nu_1+\nu_2;
\]
\[
 y_3=\eta_1-\eta_2+\frac{\gd}{1+\Psi}y_2;\hspace{0.5cm}
 y_4=\eta_1+\eta_2-\frac{\gd}{1-\Psi}y_1;
\]
\[
y_5=\xi_1-\xi_2+\frac{\gamma(1+2\gff)}{2(1-\Psi)}y_1
-\frac{2\gl+\frac{\gd}{1-\Psi}\gamma(1+2\gff)}{1-2\gff-2\frac{\gd^2}{1-\Psi}}y_4; {\normalsize\hspace{3.8cm} (A2.6)}
\]
\[
y_6=\xi_1+\xi_2+\frac{\gamma(1-2\gff)}{2(1+\Psi)}y_2+ \frac{2\gl+\gamma\gd
\frac{1-2\gff}{1+\Psi}}{1+2\gff-2\frac{\gd^2}{1+\Psi}}y_3;
\]
\[
\widetilde{\nu}=\nu-\sum_i\frac{\lan\nu y_i\ran}{\lan y_1^2\ran}y_i\ ,
\]
then $M_{1,2}$ can be written in a diagonal form. In $M_1$, $\lan
y_i^2\ran$, $i=1, ...,6$ occupy the first six positions, and
$\lan\widetilde{\nu}^2\ran$ stands on the seventh position. In matrix $M_2$,
all six diagonal positions are occupied by $\lan y_i^2\ran$,  $i=1, ...,6$.
The difference between the two quadratic
forms is  then, $Q_1-Q_2=\ha\widetilde{\nu}/\lan\widetilde{\nu}^2\ran$. Also
$\det M_1=\det M_2\cdot\lan\widetilde{\nu}^2\ran$. Eq.(A2.3) has
now a very simple form:
\[
P(\nu|B;C)d\nu(x)=\frac{1}{\sqrt{2\pi\lan\widetilde{\nu}^2\ran}}
e^{-\frac{\widetilde{\nu}^2}{2\lan\widetilde{\nu}^2\ran}} d\nu
{\normalsize\hspace{6.8cm} (A2.7)}
\]
As it is seen from (A2.6) and (A2.7), $\nu$ has the mean value:
\[
\lan\nu\ran=\sum_i\frac{\lan\nu y_i\ran}{\lan
y_i^2\ran}y_i{\normalsize\hspace{10.3cm} (A2.8)}
\]
and the variance:
\[
\lan\nu^2\ran=1-\sum_i\frac{\lan\nu y_i\ran^2}{\lan y_i^2\ran^2}y_i\
.{\normalsize\hspace{9.0cm} (A2.9)}
\]
The values in (A2.8) and (A2.9) are as follows:
\[
\lan\nu y_1\ran=\Psi_1-\Psi_2;\hspace{1cm}
\lan\nu y_2\ran=\Psi_1+\Psi_2;
\]
\[
\lan\nu y_3\ran=-\gd_1\frac{x-x_1}{r_1}+ \gd_2\frac{x-x_2}{r_2} +
\frac{\gd}{1-\Psi}\lan\nu y_2\ran;
\]
\[
\lan\nu y_4\ran=-\gd_1\frac{x-x_1}{r_1}- \gd_2\frac{x-x_2}{r_2} -
\frac{\gd}{1+\Psi}\lan\nu y_1\ran;
\]
\[
\lan\nu y_5\ran= \gamma(\gff_1-\gff_2)+\gamma\frac{1+2\gff}{2(1-\Psi)}\lan\nu
y_1\ran-\frac{2\gl+\frac{\gd}{1-\Psi}\gamma(1+2\gff)}
{1-2\gff-2\frac{\gd^2}{1-\Psi}}\lan\nu y_4\ran;
\]
\[
\lan\nu y_6\ran= \gamma(\gff_1+\gff_2)+\gamma\frac{1-2\gff}{2(1+\Psi)}\lan\nu
y_2\ran +\frac{2\gl+\frac{\gd}{1+\Psi}\gamma(1-2\gff)}
{1+2\gff-2\frac{\gd^2}{1+\Psi}}\lan\nu y_3\ran;
\]
\[
\lan y_1^2\ran=2(1-\Psi);\hspace{1cm}\lan y_2^2\ran=2(1+\Psi);
\]
\[
\lan y_3^2\ran=1+2\gff-2\frac{\gd^2}{1+\Psi};\hspace{1cm}
\lan y_4^2\ran=1-2\gff-2\frac{\gd^2}{1-\Psi};
\]
\[
\lan y_5^2\ran= \frac{3}{4}-2f-\frac{\gamma^2(1+2\gff)^2}{2(1-\Psi)}-
\frac{\left(2\gl+\frac{\gd}{1-\Psi}\gamma(1+2\gff)\right)^2}
{1-2\gff-2\frac{\gd^2}{1-\Psi}};
\]
\[
\lan y_6^2\ran= \frac{3}{4}+2f-\frac{\gamma^2(1-2\gff)^2}{2(1+\Psi)}-
\frac{\left(2\gl+\frac{\gd}{1+\Psi}\gamma(1-2\gff)\right)^2}
{1+2\gff-2\frac{\gd^2}{1+\Psi}}\ .
\]
In order to construct maxima in $x_1$ and $x_2$, it is necessary to take
into account that $\eta_{1,2}=0$ and $\xi_{1,2}<0$.

\section{Appendix B1}
In this appendix, we derive the differential and the integrated density
of the saddle points only, since the density of maxima and minima was
obtained by (B.E.). Here we present their result and compare it with
the result for the saddle points, which can be obtained in a similar
fashion.  Following (B.E.), the covariance matrix $M$ for the values
$\nu$, $\eta_1$ $\eta_2$, $\xi_{11}$,  $\xi_{22}$ and  $\xi_{12}$ can
be written in the following form:
\[
M=
\left(\begin{array}{cccccc}
1 & 0 & 0 & -\gc/2 & -\gc/2 & 0\\
0 & 1/2 & 0 & 0 & 0 & 0\\
0 & 0 & 1/2 & 0 & 0 & 0\\
-\gc/2 & 0 & 0 & 3/8 & 1/8 & 0 \\
-\gc/2 & 0 & 0 & 1/8 & 3/8 & 0\\
0 & 0 & 0 & 0 & 0 & 1/8
\end{array}
\right)\ .
 {\normalsize\hspace{4.4cm} (B1.1)}
\]
The joint probability is:
\[ P(\nu,\eta_1,\eta_2,\xi_{11},\xi_{22},\xi_{12}) d\nu d\eta_1 d\eta_2
d\xi_{11} d\xi_{22} d\xi_{12}=
\]
\[
=\frac{2}{\pi^3}\frac{1}{\sqrt{1-\gc^2}}e^{-Q}  d\nu d\eta_1 d\eta_2
d\xi_{11} d\xi_{22} d\xi_{12}{\normalsize\hspace{6.3cm} (B1.2)}
\]
\[
Q=\frac{\nu^2}{2}+\eta_1^2+\eta^2_2+(\xi_{11}-\xi_{22})^2+
\frac{(\xi_{11}+\xi_{22}+\gc\nu)^2}{2(1-\gc^2)}+4\xi_{12}^2.
\]
The differential density of extrema is:
\[
N_{ext}(\nu)=\int|\det\xi|P\cdot\gd(\eta_1) \gd(\eta_2)\cdot
d\eta_1...d\xi_{12}.
\]
We rotate the coordinate system by the angle $\gff=\ha
{\normalsize{arctg}}\frac{2\xi_{12}} {\xi_{11}-\xi_{22}}$ to align
with  the principal axes of $\xi_{ij}$. Thus, we get the diagonal form
--- $\gl_1\gl_2$, ordered by $\gl_1\ge \gl_2$, and introduce  variables
$\gl_1$, $\gl_2$, $\gff$:
\[
\begin{array}{lc}
\xi_{11}=\gl_1\cos^2\gff+\gl_2\sin^2\gff,\\
\xi_{22}=\gl_1\sin^2\gff+\gl_2\cos^2\gff,\\
\xi_{12}=(\gl_1-\gl_2)\sin\gff\cos\gff.\\
\end{array}
{\normalsize\hspace{8.1cm} (B1.3)}
\]
Thus, the transformation of the volume element is:
\[
d\xi_{11}d\xi_{22}d\xi_{12}=(\gl_1-\gl_2)d\gl_1d\gl_2d\gff.
{\normalsize\hspace{6.8cm} (B1.4)}
\]
The orientation angle $\gff$ is randomly distributed over
$0\le\gff\le\pi$. The differential density of extrema is:
\[
N_{ext}(\nu)d\nu=\frac{1}{4\pi^2}\frac{1}{\sqrt{1-\gc^2}}
\frac{\gs_2^2}{\gs^2_1} e^{-\frac{\nu^2}{2}}d\nu\int_{sad}
e^{-\frac{(4+\gc\nu)^2}{2(1-\gc^2)}}x|y^2-x^2| e^{-x^2}dxdy.
{\normalsize\hspace{0.8cm} (B1.5)}
\]
(B1.5) was obtained from (B1.1)-(B1.4) by means of substituting:
\[
\gl_1=\frac{x+y}{2}\ ,\hspace{1cm}\gl_2=\frac{y-x}{2}\ .
{\normalsize\hspace{7.5cm}(B1.6)}
\]
The limits of the integration over $S$ depends on the type of the extremal
points,
namely
\[
S:=
\left\{\begin{array}{ccc}
0>\gl_1\ge\gl_2 & - &{\normalsize{ maxima}}\\
\gl_1\ge0,\gl_2\le0 & - & {\normalsize{saddle\quad points}}\\
0<\gl_2\le\gl_2 & - &{\normalsize{ minima}}
\end{array}\right.
 {\normalsize\hspace{5.3cm} (B1.7)}
\]
If we are interested in the density of the saddle points, then
\[
N_{sad}(\nu)d\nu= {\normalsize\hspace{10.7cm} (B1.8)}
\]
\[
=\frac{1}{4\pi^2}\frac{1}{\sqrt{1-\gc^2}} \frac{\gs_2^2}{\gs^2_1}
e^{-\frac{\nu^2}{2}}d\nu\int_0^{+\infty}xe^{-x^2}dx \int_{-x}^x
e^{-\frac{(y+\gc\nu)^2}{2(1-\gc^2)}}(x^2-y^2)dy.
\]
After integrating, we get a very simple result:
\[
N_{sad}(\nu)d\nu=\frac{1}{\sqrt{32\pi}}\frac{1}{\pi}
\frac{\gs^2_1}{\gs^2_1}\frac{1}{\sqrt{3-2\gc^2}}
e^{-\frac{3\gc^2}{2(3-2\gc^2)}}d\nu\ .{\normalsize\hspace{4.7cm} (B1.9)}
\]
The integrated density is
\[
n_{sad}(\nu_t)=\int^{\infty}_{\nu_t}N_{sad}(\nu)d\nu=\frac{1}{8\pi\sqrt{3}}
\frac{\gs_2^2}{\gs_1^2}\left[1-\gF\left(\frac{\nu_t\sqrt{3}}
{\sqrt{2(3-2\gc^2)}}\right)\right],{\normalsize\hspace{1.4cm} (B1.10)}
\]
where $\gF(x)=\frac{2}{\sqrt{\pi}}\int_0^xe^{-t^2}dt$ is the probability
integral (Gradshteyn  and Ryzhik, 1980).  The density of the saddle
points of arbitrary height is
\[
n_{sad}(-\infty)=\frac{1}{4\pi\sqrt{3}}\frac{\gs^2_2}{\gs_1^2}\ .
{\normalsize\hspace{8.7cm} (B1.11)}
\]
Analogous results were obtained by (B.E.) but for differential and integrated
densities of the maxima and minima.
The differential density of maxima is
\[
N_{max}(\nu)d\nu=\frac{\sqrt{2}}{4\pi\sqrt{\pi}}
\frac{\gs_2^2}{\gs_1^2}\cdot
\left\{\frac{1}{4\sqrt{3-2\gc^2}} e^{-\frac{3\nu^2}{2(3-2\gc^2)}}
\left[1+\gF\left(\frac{\gc\nu}{\sqrt{(2-2\gc^2)(3-2\gc^2)}}\right)\right]
\right.
\]
\[
\left.+\frac{\gc(\nu^2-1)}{4}e^{-\frac{\nu^2}{2}}
\left[1+\gF\left(\frac{\gc\nu}{\sqrt{2(1-\gc^2)}}\right)\right]
+\frac{\gc\nu\sqrt{1-\gc^2}}{\sqrt{2\pi}}e^{-\frac{\nu^2}{2(1-\gc^2)}}\right\}.
{\normalsize\hspace{1.cm}(B1.12)}
\]
Integrated number density must be evaluated numerically,
\[
n_{max}(\nu_t)=\int_{\nu_t}^{\infty}N_{max}(\nu)d\nu.
{\normalsize\hspace{8.1cm}(B1.13)}
\]
The density  of maxima of arbitrary height is
\[
n_{max}(-\infty,\gc)=\frac{1}{8\pi\sqrt{3}}\frac{\gs_2^2}{\gs_1^2}\ .
{\normalsize\hspace{7.9cm}(B1.14)}
\]
Note the following important statement:
\[
n_{max}(\nu_t)+n_{min}(\nu_t)-n_{sad}(\nu_t)={\normalsize\hspace{6.8cm}(B1.15)}
\]
\[
\frac{1}{8\pi\sqrt{2\pi}}\frac{\gs^2_2}{\gs^2_1}\int_0^{+\infty}
e^{-\frac{y^2}{2}}(y^2-1)\cdot\left[ 2-\gF
\left(\frac{\nu_t+\gc\nu}{\sqrt{2(1-\gc^2)}}\right)- \gF
\left(\frac{\nu_t-\gc\nu}{\sqrt{2(1-\gc^2)}}\right)\right]dy\ .
\]
Using the property $\gF(x)=-\gF(-x)$ and (B1.15), we obtain
\[
n_{max}(0)+n_{min}(0)-n_{sad}(0)=0.{\normalsize\hspace{6.9cm}(B1.16)}
\]

\section{Appendix B2}

Similar to one-dimension, we define events $B$, $C$  as the presence of
two maxima at the points 1 and 2 respectively with parameters:
$\nu^1$;  $\eta^1_1$; $\eta^1_2$; $\xi^1_{11}$; $\xi^1_{22}$;
$\xi^1_{12}$ - for the first maximum and  $\nu^2$;  $\eta^2_1$;
$\eta^2_2$; $\xi^2_{11}$; $\xi^2_{22}$; $\xi^2_{12}$ for the second
one. We introduce the coordinate system $x,y$ as following. Let the
first maximum is located at the point $(0,0)$ and the second maximum is
located at the point $\ov{r}_0=(r_0,0)$. The
conditional probability for $\nu(\ov{r})$ to be in the range from
$\nu(\ov{r})$ to $\nu(\ov{r})+d\nu(\ov{r})$ is
\[
P(\nu|B;C)d\nu(\ov{r})=\frac{1}{\sqrt{2\pi}}\sqrt{\frac{\det M_2}{\det M_1}}
e^{-(Q_1-Q_2)}d\nu(\ov{r})\ , {\normalsize\hspace{4.cm}(B2.1)}
\]
where $M_1$ and $M_2$ are the $13\times13$ and $12\times12$ covariance matrices
respectively, $Q_1$ and $Q_2$ are the quadratic forms.
Elements of these matrices can be written in terms of the functions:
\[
\psi=\frac{c}{\gs^2_0}; \hspace{0.5cm}
\gd=\frac{c'}{\gs_1\gs_0}; \hspace{0.5cm}
\gff=\frac{c''}{\gs^2_1};\hspace{0.5cm}
\widetilde{\gff}=\frac{c_1}{\gs^2_1}; \hspace{0.5cm}
\gl=\frac{c'''}{\gs_1\gs_2}; \hspace{0.5cm}
\widetilde{\gl}=\frac{c'_1}{\gs_1\gs_2}
\]
\[
f=\frac{c''''}{\gs^2_2}; \hspace{0.5cm}
\widetilde{f}=\frac{c''}{\gs^2_2}; \hspace{0.5cm}
\widetilde{\widetilde{f}}=\frac{c_2}{\gs^2_2},
{\normalsize\hspace{6.9cm}(B2.2)}
\]
where
\[
c=\pi\int kC(k)J_0(kr)dk,
\]
is the correlation function and $c^{(n)}$ for $n=1,...,4$ its derivatives,
\[
c_1=-\pi\int k^3C(k)J_0(kr)dk,
\]
$c_1^{(n)}$ for $n=1,2$ is its derivatives, and
\[
c_2=\pi\int k^5C(k)J_0(kr)dk.
\]
The matrices $M_1$ and $M_2$ can be written in a convenient and simple diagonal
forms:
\[
M_{1ij}=\lan z_iz_j\ran\gd_{ij}, \hspace{0.5cm} i,j,=1,...,13,
\]
\[
M_{2ij}=\lan z_iz_j\ran\gd_{ij}, \hspace{0.5cm} i,j,=1,...,12.
\]
The difference between two quadratic forms is then $Q_1-Q_2=\ha
z_{13}/\lan z_{13}^2\ran$. Also $\det M_1=\lan z_{13}^2\ran\det
M_2$ and for conditional probability one can write:
\[
P(\nu|B;C)d\nu(\ov{r})=\frac{1}{\sqrt{2\pi\lan z_{13}^2\ran}}
e^{-\frac{z^2_{13}}{2\lan z_{13}^2\ran}} d\nu\ ,
{\normalsize\hspace{4.cm}(B2.3)}.
\]
The variable $z_{13}$ is the linear combination of $\nu=\nu(x,y)$ and
$z_i$, $i=1, ... ,13$:
\[
z_{13}=\nu-\sum_i\frac{\lan\nu z_i\ran}{\lan z^2_i\ran}z_i.
{\normalsize\hspace{6.9cm}(B2.4)}
\]
As it is seen from Eqs.(B2.3) and (B2.4), that $\nu$ has the mean value:
\[
\lan\nu\ran=\sum_i\frac{\lan\nu z_i\ran}{\lan z^2_i\ran}z_i,
{\normalsize\hspace{6.9cm}(B2.5)}
\]
and the variance:
\[
\lan\nu^2\ran=\sum_i\frac{\lan\nu z_i\ran^2}{\lan z^2_i\ran}.
{\normalsize\hspace{6.9cm}(B2.6)}
\]
The variables $z_i$ and $\lan z_i^2\ran$ are as follows:
\[
z_1=\nu^1-\nu^2; \hspace{0.5cm}
z_1^2=2(1-\psi); \hspace{0.5cm}
z_2=\nu_1+\nu_2; \hspace{0.5cm}
z_2^2=2(1+\psi); \hspace{0.5cm}
\]
\[
z_3=\eta^1_1-\eta_1^2+\frac{\gd}{1+\psi}z_2; \hspace{0.5cm}
\lan z_3^2\ran=1+2\gff-2\frac{\gd^2}{1+\psi}; \hspace{0.5cm}
\]
\[
z_4=\eta^1_1+\eta_1^2-\frac{\gd}{1-\psi}z_1; \hspace{0.5cm}
\lan z_4^2\ran=1-2\gff-2\frac{\gd^2}{1-\psi}
\]
\[
z_5=\eta_2^2-\eta_2^2; \hspace{0.5cm}
\lan z^2_5\ran=1-2(\gff-\widetilde{\gff});\hspace{0.5cm}
z_6=\eta_2^2+\eta_2^2; \hspace{0.5cm}
\lan z^2_6\ran=1+2(\gff-\widetilde{\gff}); \hspace{0.5cm}
\]
\[
z_7=\xi^1_{11}-\xi^2_{11}+\frac{\gc(1+2(\widetilde{\gff}-\gff))} {\lan
z_1^2\ran}z_1-\frac{2\gl+\frac{\gc\gd((1+2(\widetilde{\gff}-\gff))}{1-\psi}}
{\lan z_4^2\ran}z_4;
\]
\[
\lan z_7^2\ran=\frac{3}{4}-2f-\frac{\gc^2(1+2(\widetilde{\gff}-\gff))^2} {\lan
z_1^2\ran}-
\frac{\left(2\gl+\frac{\gc\gd((1+2(\widetilde{\gff}-\gff))}{1-\psi}\right)^2}
{\lan z_4^2\ran};
\]
\[
z_8=\xi^1_{11}+\xi^2_{11}+\frac{\gc(1-2(\widetilde{\gff}-\gff))} {\lan
z_2^2\ran}z_1+\frac{2\gl+\frac{\gc\gd((1-2(\widetilde{\gff}-\gff))}{1+\psi}}
{\lan z_3^2\ran}z_3;
\]
\[
\lan z_8^2\ran=\frac{3}{4}+2f-\frac{\gc^2(1-2(\widetilde{\gff}-\gff))^2} {\lan
z_2^2\ran}-
\frac{\left(2\gl+\frac{\gc\gd((1-2(\widetilde{\gff}-\gff))}{1+\psi}\right)^2}
{\lan z_3^2\ran};
\]
\[
z_9=\xi^1_{22}-\xi^2_{22}-\xi^1_{11}+\xi^2_{11}-
\frac{2(\widetilde{\gl}-2\gl)}{\lan z_4^2\ran}z_4-
\]
\[\left(\frac{4f-2\widetilde{f}-\ha}{\lan z_7^2\ran}-
\frac{2(\widetilde{\gl}-2\gl)(2\gl+
\frac{\gc\gd((1+2(\widetilde{\gff}-\gff))}{1-\psi}}{\lan z_4^2\ran\lan
z_7^2\ran}\right)z_7;
\]
\[
\lan z^2_9\ran=\frac{3}{4}-2(f-2\widetilde{f}+\widetilde{\widetilde{f}})-
\frac{\gc^2(1+2(\widetilde{\gff}-\gff))^2} {\lan z_1^2\ran}-
\frac{\left(2(\widetilde{\gl}-\gl)+
\frac{\gc\gd((1+2(\widetilde{\gff}-\gff))}{1-\psi}\right)^2} {\lan z_4^2\ran}-
\]
\[
\frac{\left\{\frac{1}{4}+2(f-\widetilde{f})-
\frac{\gc^2(1+2(\widetilde{\gff}-\gff))^2} {\lan z_1^2\ran} -
\frac{\left[2\gl+\frac{\gc\gd((1+2(\widetilde{\gff}-\gff))}{1-\psi}\right]
\left[2(\widetilde{\gl}-\gl)+
\frac{\gc\gd((1+2(\widetilde{\gff}-\gff))}{1-\psi}\right]} {\lan
z_4^2\ran}\right\}^2}{\lan z_7^2\ran}
\]
\[
z_{10}=\xi^1_{22}+\xi^2_{22}-\xi^1_{11}-\xi^2_{11}+
\frac{2(\widetilde{\gl}-2\gl)}{\lan z_3^2\ran}z_3-
\]
\[\left[\frac{2\widetilde{f}-4f-\ha}{\lan z_8^2\ran}-
\frac{2(\widetilde{\gl}-2\gl)(2\gl+
\frac{\gc\gd((1-2(\widetilde{\gff}-\gff))}{1+\psi}}{\lan z_3^2\ran\lan
z_8^2\ran}\right]z_8;
\]
\[
\lan z^2_{10}\ran=\frac{3}{4}+2(f-2\widetilde{f}+\widetilde{\widetilde{f}})-
\frac{\gc^2(1-2(\widetilde{\gff}-\gff))^2} {\lan z_2^2\ran}-
\frac{\left[2\gl-(\widetilde{\gl})-
\frac{\gc\gd((1-2(\widetilde{\gff}-\gff))}{1+\psi}\right]^2} {\lan z_3^2\ran}-
\]
\[
\frac{\left\{\frac{1}{4}+2(\widetilde{f}-f)-
\frac{\gc^2(1-2(\widetilde{\gff}-\gff))^2} {\lan z_1^2\ran} -
\frac{\left[2\gl+\frac{\gc\gd((1-2(\widetilde{\gff}-\gff))}{1+\psi}\right]
\left[2(\widetilde{\gl}-\gl)+
\frac{\gc\gd((1-2(\widetilde{\gff}-\gff))}{1+\psi}\right]} {\lan
z_4^2\ran}\right\}^2}{\lan z_8^2\ran};
\]
\[
z_{11}=\xi^1_{12}-\xi^2_{12}-\frac{2(\widetilde{\gl}-\gl)}{\lan z^2_6\ran}z_6;
\]
\[
\lan z_{11}^2\ran =\frac{1}{4}+2(f-\widetilde{f})-
\frac{4(\widetilde{\gl}-\gl)}{\lan z^2_6\ran}z_6;
\]
\[
z_{12}=\xi^1_{12}+\xi^2_{12}-\frac{2(\widetilde{\gl}-\gl)}{\lan z^2_5\ran}z_5;
\]
\[
\lan z_{12}^2\ran =\frac{1}{4}+2(f-\widetilde{f})-
\frac{4(\widetilde{\gl}-\gl)}{\lan z^2_5\ran}z_5.
\]
For values $\lan\nu z_i\ran$, $i=1, ..., 12$ we obtained:
\[
\lan\nu z_1\ran=\psi_1-\psi_2; \hspace{0.5cm}
\lan\nu z_2\ran=\psi_1+\psi_2;
\]
\[
\lan\nu z_3\ran=-\frac{x}{r_1}\gd_1+\frac{x-r_0}{r_2}\gd_2+\frac{\gd}{1+\psi}
\lan\nu z_2\ran;
\]
\[
\lan\nu z_4\ran=-\frac{x}{r_1}\gd_1-\frac{x-r_0}{r_2}\gd_2-\frac{\gd}{1-\psi}
\lan\nu z_1\ran;
\]
\[
\lan\nu z_5\ran=-\frac{y}{r_1}\gd_1+\frac{y}{r_2}\gd_2; \hspace{0.5cm}
\lan\nu z_6\ran=-\frac{y}{r_1}\gd_1-\frac{y}{r_2}\gd_2;
\]
\[
\lan\nu z_7\ran=\frac{x^2-y^2}{r_2^2}\gff_1-\frac{(x-r_0)^2-y^2}{r^2_2}\gff_2+
\frac{y^2}{r_1^2}\widetilde{\gff}_1-\frac{y^2}{r_2^2}\widetilde{\gff}_2+
\]
\[
\frac{\gc(1+2(\widetilde{\gff}-\gff))}{\lan z_1^2\ran}\lan\nu z_1\ran-
\frac{2\gl+\frac{\gc\gd((1+2(\widetilde{\gff}-\gff))}{1-\psi}} {\lan
z_4^2\ran}\lan\nu z_4\ran;
\]
\[
\lan\nu z_8\ran=\frac{x^2-y^2}{r_2^2}\gff_1-\frac{(x-r_0)^2-y^2}{r^2_2}\gff_2+
\frac{y^2}{r_1^2}\widetilde{\gff}_1+\frac{y^2}{r_2^2}\widetilde{\gff}_1+
\]
\[
\frac{\gc(1-2(\widetilde{\gff}-\gff))}{\lan z_2^2\ran}\lan\nu z_2\ran +
\frac{\left[2\gl+\frac{\gc\gd((1+2(\widetilde{\gff}-\gff))}{1-\psi}\right]}
{\lan z_3^2\ran}\lan\nu z_3\ran;
\]
\[
\lan \nu z_9\ran=\frac{y^2-x^2}{r_1^2}(2\gff_1-\widetilde{\gff}_1)-
\frac{y^2-(x-r_0)^2}{r_2^2}(2\gff_2-\widetilde{\gff}_2)-
\frac{2(\widetilde{\gl}-2\gl)}{\lan z_4^2\ran}\lan\nu z_4\ran-
\]
\[
\left[\frac{4f-2\widetilde{f}-\ha}{\lan z_7^2\ran}-
\frac{2(\widetilde{\gl}-2\gl)(2\gl+
\frac{\gc\gd((1+2(\widetilde{\gff}-\gff))}{1-\psi}}{\lan z_4^2\ran\lan
z_7^2\ran}\right]\lan\nu z_7\ran;
\]
\[
\lan \nu z_{10}\ran=\frac{y^2-x^2}{r_1^2}(2\gff_1-\widetilde{\gff}_1)+
\frac{y^2-(x-r_0)^2}{r_2^2}(2\gff_2-\widetilde{\gff}_2)+
\frac{2(\widetilde{\gl}-2\gl)}{\lan z_4^2\ran}\lan\nu z_4\ran-
\]
\[
\left[\frac{2\widetilde{f}-4f-\ha}{\lan z_8^2\ran}-
\frac{2(\widetilde{\gl}-2\gl)(2\gl+
\frac{\gc\gd((1+2(\widetilde{\gff}-\gff))}{1+\psi}}{\lan z_3^2\ran\lan
z_8^2\ran}\right]\lan\nu z_8\ran;
\]
\[
\lan \nu z_{11}\ran=\frac{xy}{r_1^2}(2\gff_1-\widetilde{\gff}_1)-
\frac{(x-r_0)y}{r_2^2}(2\gff_2-\widetilde{\gff}_2)-
\frac{2(\widetilde{\gl}-\gl)}{\lan z_6^2\ran}\lan\nu z_6\ran;
\]
\[
\lan \nu z_{12}\ran=\frac{xy}{r_1^2}(2\gff_1-\widetilde{\gff}_1)+
\frac{(x-r_0)y}{r_2^2}(2\gff_2-\widetilde{\gff}_2)-
\frac{2(\widetilde{\gl}-\gl)}{\lan z_5^2\ran}\lan\nu z_5\ran;
\]

\newpage
\begin{center}
References
\end{center}

\noi
Bardeen, J.M., Bond, J.R, Kaiser, N. \&  Szalay, A.S. 1986, ApJ, {\bf 304}, 15

\noi
Bond, J.R. \& Efstathiou, G. 1987, MNRAS, {\bf 226}, 655

\noi
Cheng, E.S. et al. 1994, ApJ, {\bf 422}, L37

\noi
Clapp, A.C. et al. 1994, ApJ, {\bf 433}, L57

\noi
Coulson,D., Ferreira, P., Graham, P. \& Turok, N. 1994, Nature, {\bf 368}, 27

\noi
De Bernardis P. et al. 1994, ApJ, {\bf 422}, L33

\noi
Devlin, M.J. et al. 1994, ApJ, {\bf 422}, L1

\noi
Dominik, K. \& Shandarin, S. 1992, ApJ, {\bf 393}, 450

\noi
Doroshkevich, A.G. 1970, Astrophysica, {\bf 6}, 320

\noi
Dragovan, M., Ruhl, J.E., Novak, G., Platt, S.R., Crone, B., Pernic, R. \&
Peterson, J.B. 1994, ApJ, {\bf 427}, L67

\noi
Efstathiou, G., Bond, J.H.R., \& White, S.D.M. 1992, MNRAS, {\bf 258}, 1P

\noi
Einasto, J., Klypin, A.A., Saar, E. \& Shandarin, S. 1984, MNRAS, {\bf 206},
529

\noi
Gorski, K. 1993, ApJ, {\bf 410}, L65

\noi
Gradshteyn, I.S. \& Ryzhik, I.M. 1980, Tables of Integrals, Series and
Products. Ed. by Alan Jeffrey. Academic Press.

\noi
Hinshaw, G., Bennet, C. \& Kogut, A. 1995, ApJ, {\bf441} L1

\noi
J\o rgensen, H., Kotok, E., Naselsky, P. \& Novikov, I. 1995, Astron.
Astrophys, {\bf 294}, 639

\noi
Klypin, A.A. 1985, Soviet Astron. {\bf 31}, 8

\noi
Naselsky, P., \& Novikov, I. 1993; ApJ, {\bf 413}, 14

\noi
Rice, S.O. 1944, Bell System Tech.J., {\bf 23}, 282

\noi
Rice, S.O. 1945, Bell System Tech.J., {\bf 24}, 41

\noi
Schuster, J., et al. 1993, ApJ, {\bf 412}, L47

\noi
Scott, D., Silk, J. \& White, M. 1995, Preprint

\noi
Starobinsky, A.A. 1992, JETP Lett., {\bf 55}, 489

\noi
Wollack, E.J., Jarosik, N.C., Netherfeld, C.B., Page, L.A.S. \& Wilkinson, D.
1994, ApJ, {\bf 419}, L49

\noi
Zel'dovich, Ya.B. 1982, Soviet Astron Lett, {\bf 8}, 102

\newpage
\begin{center}
{\bf Figure captions}
\end{center}

{\bf Fig.1} Correlation function for CDM models with Harrison-Zel'dovich
spectrum of the initial perturbations, $\gW_0=1$, $h=0.5$.
(a) Filtered correlation function with filters $H_0$ and $H_2$. Dashed
line: $\gW_b=0.1$. Dotted line: $\gW_b=0.03$. Solid line: models
without a Doppler-peak and $C_l=\frac{1}{l(l+1)}$.
(b) The same as Fig.1a, but for filtered correlation functions with filter
$H_*(\gd_{max},\gd_{min}$,r).
1: for $\gd_{max}\sim l=10$, $\gd_{min}\sim l=150$;
2: for $\gd_{max}\sim l=30$, $\gd_{min}\sim l=150$.
(c) Filtered correlation functions for the filter $H_2$ with the
following width $\gd$ (curves from the left to the right):(1) $\gd\sim
l=660$, (2) $\gd\sim l=350$, (3) $\gd=0.67^o\sim l=150$ corresponding to the
resonance filter, (4) $\gd\sim l=100$, (5) $\gd\sim l=75$, (6) $\gd\sim
l=50$.

{\bf Fig.2} Correlation functions for the model $\gW_b=0.03$ (ensemble average)
shown as solid lines and their dispersions $C(r)\pm \sqrt{D(r)}$ as dashed
lines; realizations are shown as *. The field size is $10^o\times10^o$. The
upper
part of the figure, marked by $H_o$,
corresponds to the unfiltered Map 1a. The lower part, marked by $H_2$,
corresponds
to the filtered Map 1b.

{\bf Fig.3} The dependence of the spectral parameter $\gc$ on the value of the
filter width $\gd$.
(a) For the filters $H_0$ and $H_2$. The solid line is without a Doppler-peak;
the dotted line corresponds to $\gW_b=0.03$; the  dashed line to $\gW_b=0.1$.
(b) The same as Fig.3a, but for the filter $H_*$. 1: corresponds to
$\gd_{max}\sim l=10$,  2: corresponds to $\gd_{max}\sim l=30$.

{\bf Fig.4} (a) A one-dimensional cut of the simulated map $10^o\times 10^o$.
Arrows indicate two maxima, which are tested in Fig.4b.
(b) The solid line is an expected value of the field in the vicinity of two
neighbouring maxima indicated in Fig.4a. The dashed lines correspond to
expected values $\pm\sqrt{variance}$.

{\bf Fig.5} Clusterisation of Maxima in one-dimension for different values of
$\gc$.
(a) Differential densities of maxima and minima, solid line for $\gc=0.1$,
dashed line for $\gc=0.5$, dotted line for $\gc=0.9$.
(b) the mean length of the clusters: solid line for $\gc=0.1$, dashed line for
$\gc=0.5$, dotted line for $\gc=0.9$.

{\bf Fig.6} $\gD T/T$ as a two-dimensional surface in a three-dimensional
space.

{\bf Fig.7} Clusterisation of peaks in the two-dimensional case.
(a) Differential density of maxima shown as a dotted line; saddle points as
a solid line; minima as a  dashed line.
(b) Mean length of clusters of maxima (right-hand side) and minima (left-hand
side) for different $\gc$: solid line for $\gc=0.1$, dashed line for $\gc=0.5$,
dotted line for $\gc=0.9$.

{\bf Fig.8} Different slices of the $\gD T/T$ maps in the vicinity of two
maxima. (a) a model with $\gW_b=0.1$, (b) a model without a Doppler-peak.

{\bf Map 1} (a) Simulated map of $\gD T/T$ for a  region of $10^o\times 10^o$
for a  model with $\gW_b=0.03$.
(b) The same as for (a) but with the filter $H_2$, $\gd=0.67^o$.

{\bf Map 2} Simulated map of $\gD T/T$ for a region of $10^o\times 10^o$,
for a model with $\gW_b=0.03$ and filter $H_2$, $\gd=0.67^o$. *=maxima,
X=saddle
 points;
contours correspond to the level $\nu_t=1$.

{\bf Map 3} Percolation through the black zone ($\nu_t>0$).

{\bf Map 4} Simulated maps  of $10^o\times 10^o$ for two different models
(see text)  (a) $\gW_b=0.1$, (b) Model without a Doppler-peak.

\end{document}